%% file: 01-main.tex
\documentclass[10pt,conference]{IEEEtran}

\def\BibTeX{{\rm B\kern-.05em{\sc i\kern-.025em b}\kern-.08emT\kern-.1667em\lower.7ex\hbox{E}\kern-.125emX}}

\usepackage{setspace}
\usepackage{bm}
\usepackage{xcolor}
\usepackage{booktabs}
\usepackage{microtype}
\usepackage{amsfonts}
\usepackage{xspace}

\usepackage{enumitem,etoolbox}
\usepackage{array}
\usepackage{microtype}
\usepackage{multirow}
\usepackage{framed}
\usepackage{caption}
\usepackage{threeparttable}
\usepackage{fontawesome}
\usepackage[labelformat=simple]{subcaption}

\usepackage[ruled,vlined]{algorithm2e}

\SetCommentSty{mycommfont}

\SetKwInOut{Input}{Input}
\SetKwInOut{Output}{Output}
\SetKwInOut{Except}{Except.}
\SetKw{None}{nil}
\SetKw{Cont}{continue}
\SetKw{Break}{break}
\SetKwProg{Fn}{Function}{:}{}
\SetKwFunction{Initp}{InitPattern}
\SetKwFunction{Nextp}{NextPattern}
\SetKwFunction{Initss}{InitSerSched}
\SetKwFunction{Nextss}{NextSerSched}
\SetAlCapFnt{\small}
\SetAlCapNameFnt{\small}
\LinesNumbered
\DontPrintSemicolon

\usepackage{tikz}
\usetikzlibrary{positioning}
\tikzstyle{period} = [draw=white, fill=gray!30, thick,
    rectangle, rounded corners, minimum size=9ex]
\usepackage{subcaption}
\usepackage{soul}

\AtBeginEnvironment{verbatim}{\setlist[trivlist]{nosep}}
\usepackage{flushend}
\usepackage{balance}
\usepackage{graphicx}
\usepackage{amsthm}
\usepackage{amsmath}
\usepackage{amssymb}

\usepackage{bbding}
\usepackage{array}
\usepackage{multirow}
\usepackage{listings}
\usepackage{color}
\usepackage{tcolorbox}
\usepackage{xcolor,colortbl}
\usepackage[hidelinks]{hyperref}
\usepackage{enumitem}
\usepackage{adjustbox}
\usepackage[normalem]{ulem}
\usepackage{pifont}
\usepackage{wasysym}

\usepackage[hang,flushmargin]{footmisc}

\definecolor{nord0}{HTML}{2E3440}
\definecolor{nord1}{HTML}{3B4252}
\definecolor{nord2}{HTML}{434C5E}
\definecolor{nord3}{HTML}{4C566A}
\definecolor{nord4}{HTML}{D8DEE9}
\definecolor{nord5}{HTML}{E5E9F0}
\definecolor{nord6}{HTML}{ECEFF4}
\definecolor{nord7}{HTML}{8FBCBB}
\definecolor{nord8}{HTML}{88C0D0}
\definecolor{nord9}{HTML}{81A1C1}
\definecolor{nord10}{HTML}{5E81AC}
\definecolor{nord11}{HTML}{BF616A}
\definecolor{nord12}{HTML}{D08770}
\definecolor{nord13}{HTML}{EBCB8B}
\definecolor{nord14}{HTML}{A3BE8C}
\definecolor{nord15}{HTML}{B48EAD}
\definecolor{seen}{HTML}{C8D2EB}
\definecolor{unseen}{HTML}{FDF0CE}

\definecolor{downcolor}{HTML}{5573A5}
\definecolor{upcolor}{HTML}{B85B59}

\usetikzlibrary{shapes,positioning}
\makeatletter
\let\old@lstKV@SwitchCases\lstKV@SwitchCases
\def\lstKV@SwitchCases#1#2#3{}
\makeatother
\usepackage{lstlinebgrd}
\makeatletter
\let\lstKV@SwitchCases\old@lstKV@SwitchCases

\lst@Key{numbers}{none}{%
    \def\lst@PlaceNumber{\lst@linebgrd}%
    \lstKV@SwitchCases{#1}%
    {none:\\%
        left:\def\lst@PlaceNumber{\llap{\normalfont
                \lst@numberstyle{\thelstnumber}\kern\lst@numbersep}\lst@linebgrd}\\%
        right:\def\lst@PlaceNumber{\rlap{\normalfont
                \kern\linewidth \kern\lst@numbersep
                \lst@numberstyle{\thelstnumber}}\lst@linebgrd}%
    }{\PackageError{Listings}{Numbers #1 unknown}\@ehc}}
\makeatother

\definecolor{rowcolor}{HTML}{ECEFF4}

\setlist[itemize]{leftmargin=*}

\definecolor{dkgreen}{rgb}{0,0.6,0}
\definecolor{gray}{rgb}{0.5,0.5,0.5}
\definecolor{mauve}{rgb}{0.58,0,0.82}

\definecolor{bg}{HTML}{F8F9FB}  
\definecolor{bgc}{HTML}{FCF6E4}
\usepackage[framemethod=TikZ]{mdframed}

\mdfdefinestyle{MyFrame}{%
    linecolor=black,
    outerlinewidth=0.3pt,
    roundcorner=5pt,
    skipabove = 5.5pt,
    skipbelow = 5.5pt,
    innertopmargin=5.5pt, 
    innerbottommargin=5.5pt, 
    innerrightmargin=5.5pt,
    innerleftmargin=5.5pt,
    backgroundcolor=bg, 
    }

\newcommand{\etal}{\hbox{\emph{et al.}}\xspace}
\newcommand{\eg}{\hbox{\emph{e.g.}}\xspace}
\newcommand{\ie}{\hbox{\emph{i.e.}}\xspace}
\newcommand{\etc}{\hbox{\emph{etc.}}\xspace}

\setlist[itemize]{leftmargin=*}
\setlist[enumerate]{leftmargin=*}
\newlist{steps}{enumerate}{1}
\setlist[steps, 1]{label = \textbf{RQ\arabic*.}}

\newcommand{\upp}[1]{\textcolor{upcolor}{\textbf{#1}}}
\newcommand{\down}[1]{\textcolor{downcolor}{\textbf{#1}}}
\newcommand{\equ}[1]{\textcolor{nord7}{\textbf{#1}}}

\begin{document}

\title{
Concerned with Data Contamination?
Assessing Countermeasures in Code Language Model
}

\makeatletter
\newcommand{\linebreakand}{%
  \end{@IEEEauthorhalign}
  \hfill\mbox{}\par
  \mbox{}\hfill\begin{@IEEEauthorhalign}
}
\makeatother

\author{
  \IEEEauthorblockN{Jialun Cao}
  \IEEEauthorblockA{{The Hong Kong University of} \\
    \textit{Science and Technology}\\
    Hong Kong, China \\
    jcaoap@cse.ust.hk}
  \and
  \IEEEauthorblockN{Wuqi Zhang}
  \IEEEauthorblockA{{The Hong Kong University of} \\
    \textit{Science and Technology}\\
    Hong Kong, China \\
    wzhangcb@cse.ust.hk}
  \and
  \IEEEauthorblockN{Shing-Chi Cheung}
  \IEEEauthorblockA{{The Hong Kong University of} \\
    \textit{Science and Technology}\\
    Hong Kong, China \\
    scc@cse.ust.hk}
}

\maketitle

\thispagestyle{plain}
\pagestyle{plain}

\input{Tex/00-abstract}


\input{Tex/02-introduction}

\input{Tex/03-design}

\input{Tex/04-preperation}

\input{Tex/05-evaluation-rq1}

\input{Tex/05-evaluation-rq2}

\input{Tex/05-evaluation-rq3}
\input{Tex/05-evaluation-rq4}

\input{Tex/055-threat}

\input{Tex/06-relatedwork}

\input{Tex/07-conclusion}

\clearpage

\balance
\bibliographystyle{IEEEtran}
\bibliography{Tex/reference}

\end{document}

%% file: Tex/00-abstract.tex
\begin{abstract}





Various techniques have been proposed to leverage the capabilities of code language models (CLMs) for software engineering tasks. 
While these techniques typically evaluate their effectiveness using publicly available datasets, the evaluation can be subject to data contamination threats where the evaluation datasets have already been used to train the concerned CLMs. 
This can significantly affect the reliability of the evaluation. 
Different countermeasures have been suggested to mitigate the data contamination threat. 
Countermeasures include using more recent data, curating new data, and refactoring existing data are introduced, yet it is unclear whether these countermeasures could really mitigate data contamination threats to model evaluation.
To fill the gap, we systematically study to quantify the impacts of these countermeasures on CLMs' performance.

To facilitate the study, we collected 2,493,174 (over 2 million) Python functions with timestamps ranging from January 1st, 2018, to December 31st, 2023. 
The data created before the models' cut-off date are considered ``contaminated data'', while the data where the countermeasures are taken are regarded as ``cleansed data''. 
We study the impact of these countermeasures by investigating the difference in CLMs' performance on contaminated and cleansed data derived from different countermeasures.  
Our experiments yield several interesting observations. 
For instance, CLMs do not necessarily perform worse on data after the models' cut-off date; on the contrary, they sometimes perform better. In addition, refactoring did not always result in decreased performance; it could lead to improvements instead. Furthermore, existing metrics such as perplexity are incapable of distinguishing contaminated/cleansed data. 
We hope that the results and observations could help deepen the understanding of CLMs' capabilities and inform the community about data contamination when validating the effectiveness of methods.

\end{abstract}

%% file: Tex/02-introduction.tex
\section{Introduction}\label{sec:intro}

Large language models (LLMs) are increasingly utilized by state-of-the-art techniques to solve challenging software engineering problems~\cite{deng2023large,deng2024large,xia2024fuzz4all,xia2023automated}. 
Since such models were trained on a large amount of data from various code corpus, 
{{it is hard to tell whether the evaluation results of these techniques overfit the training data.
}} 
In other words, 
{{the data used for evaluation may have already been inadvertently seen}} in the training data, resulting in exaggerated performance.
The situation is identified as \textbf{\textit{data contamination}}~\cite{dataContamination2023,sainz-etal-2023-nlp-contam,taskContamination2023,li2023open}~\footnote{Other terms like ``data leakage''~\cite{VJBench23,li2023nuances,cao2023study,fan2023large}, ``memorization''~\cite{tirumala2022memorization}, ``task contamination''~\cite{taskContamination2023} are used in related works. This paper uses ``data contamination'' to represent uniformly.}.

\textbf{\textit{Challenge --}}
Identifying the existence of data contamination is difficult. In traditional machine learning scenario where the dataset scale is not that large, the contamination could be avoid by splitting training/validation/testing sets and avoid leaking target variables during training. 
However, in the era of LLMs, where models are trained on vast corpora of data spanning a diverse range of topics and sources, \textbf{\textit{avoiding the data contamination becomes expensive and sometimes unrealistic}}. First, \textbf{\textit{Opacity of training source}}. Model developers are \textbf{\textit{unwilling to disclose}} the training data and implementation details~\cite{topkmia2023,extracting21,touvron2023llama} due to various reasons (\ie, intellectual property, data privacy, license, and commercial competition), leaving no clues to the training data. Second, \textbf{\textit{Indirect contamination}}. The data pipelines for LLMs are much more complex. The preprocessing steps, such as tokenization and data augmentation, can introduce contamination indirectly. 
Third, \textbf{\textit{Scale of training data}}. Even if the training source is disclosed, the sheer volume of data makes it impractical to traverse the entire training set. Moreover, simply scanning the training data may not be reliable due to the differences in data format (\eg, XML, CSV) and their preprocessing (\eg, tokenization, and normalization).

\textbf{\textit{Research gap --}} 
Given the uncontrollability of training process of LLMs, researchers shifted to proposing \textbf{{countermeasures}} to ensure the reliability of evaluation. Various countermeasures are introduced. For example, \ding{202} Using \textbf{\textit{recent data}} (\ie, {{the data released after the deployed model's training cut-off date}}) will be less subject to contamination. These data are less likely to be used for model training. \ding{203} The community keeps releasing new calibrated-crafted code benchmarks ~\cite{humaneval,mpbb2021,du2023classeval,yu2023codereval,Lai2023DS1000,jimenez2024swebench} for fairer and more comprehensive evaluation of model capabilities. 
\ding{204} Various \textbf{\textit{code refactoring operations}}~\cite{VJBench23,shirafuji2023refactoring} (\eg, renaming the identifiers or adjusting code structures) are adopted to mutate the code before the cut-off date to avoid LLMs' memorization~\cite{VJBench23}. 
While these initiatives are promising, there is \textbf{\textit{no systematic, quantitative study to assess the impact of these countermeasures}}.

To bridge the gap, we conducted a systematic study assessing the efficacy of three countermeasures on code language models (CLMs). 
We focus on CLMs rather than general large language models due to their \textbf{\textit{relevancy}} to software engineering tasks. 
To determine whether data is {{contaminated}} or not, we use \textbf{\textit{time}} as a subjective criterion. Such {{temporal separation}} is common in time-sensitive domains like financial forecasting~\cite{huang2023finbert,gruver2024large}. The {{data created {before} the models' cut-off date}} (\ie, the date when the data collection for training concluded) are considered as ``\textbf{\textit{contaminated data}}'', where the cut-off dates vary from models. On the contrary, the data where the \textbf{\textit{countermeasures}} are adopted are regarded as ``\textbf{\textit{cleansed data}}''. 
The efficacy of countermeasures then is assessed by investigating the difference in CLMs' performance on contaminated/cleansed data, in line with related works~\cite{dataContamination2023,extracting21}. 

To facilitate the study, we collected 2 million Python codes within six years (January 1st, 2018, to December 31st, 2023). 
The \textbf{\textit{contaminated data}} is a form of the codes whose timestamps precede models' cut-off dates. The \textbf{\textit{cleansed data}} are derived from implementing countermeasures \ding{202} $\sim$ \ding{204}. 
Furthermore, we use three code similarities (\eg, Jaccard distance, Levenshtein distance as used in related work~\cite{thestackv2}) to {{quantify the overlap degree between contaminated/cleansed data}}, showing that the \textbf{\textit{overlap is at a relatively minimal level}} (Section~\ref{sec:assumption}).
In addition, we compute the code similarities between contaminated and cleansed data to make sure that the overlap between them is under a relatively small level.
We study three research questions (RQs): 
\begin{itemize}
    \item \textbf{RQ1. How does CLMs' performance differ on contaminated / \underline{\textbf{\textit{recent}}} data} (Countermeasure \ding{202})? We investigate the impact of recent data (created after models' cut-off date) as a countermeasure. We compare the CLMs' performance on recent data compared with that on contaminated data, seeing how the performance changes over time.

    \item \textbf{RQ2. How does CLMs' performance differ on contaminated / \underline{\textbf{\textit{curated}}} data} (Countermeasure \ding{203})?
    We also study whether curate data could serve as an effective countermeasure. In particular, we consider coding benchmarks HumanEval and CoderEval because of their popularity.
    Then, we analyzed the CLMs' performance differences in contaminated and curated data.
    
     \item \textbf{RQ3. How does CLMs' performance differ on contaminated / \underline{\textbf{\textit{refactored}}} data} (Countermeasure \ding{204})?
     There are several commonly adopted code refactoring operators whose impacts on CLMs' performance are worth exploring. We then apply five operators to contaminated data and see how CLMs reacts to the refactored data. 
\end{itemize}

\noindent Finally, related tasks such as membership inference attacks (MIA)~\cite{topkmia2023,loss-mia} and privacy extraction~\cite{kandpal2022deduplicating} introduce various metrics such as perplexity~\cite{jelinek1977perplexity}. While these metrics have been proven effective in related tasks, it is still unknown whether they are effective in coding tasks and CLMs. So, we further designed RQ4 to investigate MIA-related metrics for distinguishing contaminated/cleansed data.
\begin{itemize}
    \item \textbf{RQ4. Could existing metrics distinguish contaminated/cleansed data?} 
    We investigate six related metrics on contaminated and three kinds of cleansed data, 
    checking whether they are useful in distinguishing contaminated/cleansed data.
\end{itemize}

\noindent\textbf{\textit{Findings --}}
Our study yields several interesting findings. 
\begin{itemize}
    \item CLMs, in general, \textbf{\textit{perform better on countermeasure-applied data}} (\eg,~recent data, curated data, and syntactically refactored data) compared with the performance on contaminated data, indicating that the current \textbf{\textit{countermeasures may not be effective}} in alleviating data contamination.
    \item Existing MIA-related metrics, such as Perplexity, Zlib Compression Entropy, and MIN-K\% PROB can \textbf{\textit{hardly distinguish}} the contaminated/cleansed data.
    \item The popularity of AI programming assistants such as Copilot may further exacerbate data contamination in the evaluation.
     \item \textbf{\textit{Changing code structures}} may not be useful to alleviate data contamination. On the contrary, refactoring the code structure could even upgrade the models' performance. 
    \item \textbf{\textit{Semantic refactoring operators }}such as identifier renaming and appending special parameters have a greater impact on data and may be more useful to alleviate data contamination during evaluation. 
\end{itemize}

\noindent\textbf{\textit{Contributions --}} 
Our contribution is summarized as follows. 
\begin{itemize}
    \item \textbf{Novelty} We present the first study of the mitigation effect of various countermeasures for CLM data contamination. In particular, we group Python codes according to how they are collected/curated (\eg, crawled from online resources, manually crafted, or refactored from other codes) and compare how CLMs perform in these code groups. 
    
    \item \textbf{Significance} We decompose the study on the data contamination problem into four-fold, allowing for {{a more holistic analysis}}. Existing work either directly builds new datasets or applies automatic or manual code refactoring without systematically and quantitatively studying whether doing so can mitigate the threat of data contamination. 
    
    \item \textbf{Impact.}
    We collected 2 million Python codes within six years (January 1st, 2018, to December 31st, 2023), then sampled 384 Python functions each year, with a sampling confidence level of 95\%, culminating in 2304 Python functions. It serves not only as the foundation for our rigorous evaluation but also as a {{dataset for future research endeavors}}.
    
\end{itemize}

%% file: Tex/03-design.tex
\section{Experiment Design}\label{sec:design}

\input{FigureTex/figure-rq123-design}

\subsection{Experiment Design}

To facilitate the study, we construct several groups of code (\textbf{code groups}) into contaminated and cleansed data. Comparing the CLM's performance on the \textit{\textbf{contaminated}} and \textbf{\textit{cleansed}} data can infer the extent to which the CLMs rely on memorization versus their genuine ability and thus act as \textbf{\textit{a hint for the degree of data contamination}} during the evaluation.
Fig.~\ref{fig:rq123-design} visualizes an overview design for RQ1-RQ3. 
The \textbf{\textit{cleansed data}} (highlighted in \colorbox[HTML]{FFEDC1}{yellow}) could be collected after model release (RQ1), curated manually (RQ2), or mutated from contaminated data (RQ3). 
These three RQs study how CLMs' performance changes over contaminated/cleansed data.
Finally, RQ4 explores whether there are metrics that could distinguish them. 

\subsubsection{\textbf{Experiment Design for RQ1}}
In Fig.~\ref{fig:rq123-design} RQ1, codes could be partitioned into distinct code groups based on their creation \textbf{\textit{times}} (\eg, commit time in Github) because time serves as an objective indicator~\cite{huang2023finbert,gruver2024large,duan2024membership}. 
This \textbf{\textit{chronological grouping}} aims to separate the code that a CLM could have encountered during its training (\textbf{\textit{contaminated data}}) from the code generated after its training was completed (\textbf{\textit{recent data}}). 

For clarity, we designate code within a specific year with corresponding labels, such as {{Code-2018}} for snippets that fall in 2018 and similarly for other annual collections.
Note that the pivot time of contaminated/recent data varies from CLMs (See Section~\ref{sec:model}).

After the temporal splitting, we apply CLMs to these code groups, allowing us to systematically investigate how CLMs' performance changes chronologically, implying the degree of data contamination within them.

\subsubsection{\textbf{Experiment Design for RQ2}}
\label{sec:design:rq2}
Researchers and developers introduced new datasets~\cite{humaneval,yu2023codereval,du2023classeval,Lai2023DS1000,mbpp2021,jimenez2024swebench} to fairly and comprehensively evaluate the effectiveness of CLMs. 
We investigate the performance of CLMs on them.

Among the newly proposed datasets, we select two representative and popular-used function-level Python coding datasets, \ie, HumanEval~\cite{humaneval} and CoderEval~\cite{yu2023codereval}, which we refer to as {Code-HumE} and {Code-CodE}, respectively. 
We focus on function-level programs to eliminate the influence of extra-long contexts~\cite{du2023classeval}. 
In addition, since Python is the most well-trained programming language, if there are any data contamination issues, they should be more evident in Python. 
Also, most datasets and methods based on CLMs are designed specifically for Python. 
Thus, investigating data contamination threats
in Python may hold broader significance than studying in other programming languages. As shown in Fig.~\ref{fig:rq123-design} RQ2, we demonstrate the CLMs' performance in these two datasets and compare them with that in contaminated data. 

The release time of HumanEval and CoderEval are Jul 8, 2021~\cite{humanevaldata} and Jan 28, 2023~\cite{codereval4python}, respectively, according to the commit date shown in Github. 
We then compare these commit dates with models' cut-off dates to determine whether these datasets contain contaminated data regarding a CLM.

\subsubsection{\textbf{Experiment Design for RQ3}}
As proposed by existing works~\cite{VJBench23}, a more applicable way to alleviate the threat of data contamination in evaluating CLM-based approaches is to refactor the contaminated code. 
If CLMs maintain performance on refactored code, this would imply that they truly grasp the coding tasks rather than relying on rote memorization.
In contrast, if there is a substantial fluctuation in CLMs' performance between contaminated and refactored data, it indicates that CLMs are likely threatened by data contamination via memorizing specific code patterns or formats in the training data. 
Conversely, if a refactoring operator can induce significant variability (primarily decreases) in CLMs' performance, it may mitigate the CLM's exposure to data contamination.

As shown in Fig.~\ref{fig:rq123-design} RQ3, we adopted two \textbf{\textit{syntactic}} and three \textbf{\textit{semantic code refactoring operations}} (See Section~\ref{sec:refactor-ops}) to contaminated code. 
In principle, we choose the operators that could be \textbf{\textit{automated}} to avoid subjectivity brought by manual operation, reduce manual effort, and increase the \textbf{\textit{practicality and reusability}} of these refactor operators. 

On the one hand, \textbf{\textit{syntactic operators}} 
check whether CLMs rely on rote memorization of pattern recognition. 
The refactoring preserves the semantic purpose but alters its syntactic structure~\cite{baqais2020automatic,al2017empirical}. If CLMs truly understand the semantics of the code, their performance should not significantly decline when faced with semantically equivalent but syntactically varied code snippets.
On the other hand, the \textbf{\textit{semantic operators}} ensure CMLs understand code semantics rather than keyword/string matching. Replacing with synonyms or adding more semantics should not significantly affect model performance. 

The experiment design of RQ3 aims to identify refactoring operators that are more likely to \textbf{\textit{disrupt the memorized patterns}}.
The performance of CLMs on each refactoring operator is compared to assess which operators cause the greatest decline in performance, hence indicating a higher effectiveness in alleviating data contamination threats.

\subsubsection{\textbf{Experiment Design on the \underline{Assumption} of RQ1-RQ3}}
Recap that using CLMs' performance as a hint for data contamination has an assumption, \ie, \textbf{\textit{the contaminated/cleansed data are on similar difficulty level}}. 
If some code groups have a significantly higher or lower complexity than others, it is hard to explain CLMs' differential performance on code groups, obscuring true data contamination effects.
Therefore, we quantify the code complexity of different code groups used in RQ1-RQ3 using two code complexity metrics (see Section~\ref{sec:complexity}). 
\looseness=-1

Besides, we also compared the similarity between code groups horizontally using various code similarities covering syntactic and semantic information (see Section~\ref{sec:simi}), to check if any code group is significantly different from other code groups. As illustrated in Fig.~\ref{fig:rq-sim}, the average code similarity between two code groups is calculated pairwise. 

\input{FigureTex/figure-sim-design}

\subsubsection{\textbf{Experiment Design for RQ4}}
Various techniques and algorithms have been proposed for membership inference attack (MIA)~\cite{topkmia2023,carlini2019secret} or indicate how likely the model is to predict the given sequence~\cite{jelinek1977perplexity,extracting21,li2023estimating}.
Though these metrics are not designed for measuring data contamination, they could still serve as indicators. For example, perplexity indicates how well the given data fits the models' training data. 
In the study, we adopt three types of MIA-related metrics (see Section~\ref{sec:leakage-metrics}) with different parameters, resulting in six MIA-related metrics. Under each metric, we compare the scores of different code groups (\eg, contaminated/recent, contaminated/refactored), investigating whether these MIA-related metrics could distinguish the code groups. 
\looseness=-1

In addition, though these metrics could not directly serve as data contamination measurements, they could still indicate \textbf{\textit{how close the data distribution is}} between the code group and the model's training data. In particular, if a code group achieves an average lower score than other groups, it means that the distribution of the code group is closer to that of training data.

\subsection{\textbf{Coding Task}}
We chose \textit{\textbf{code completion}} because it
aligns well with the characteristics of CLMs (\ie, generative models that receive prefixes and complete suffixes)~\cite{raychev2014code,svyatkovskiy2019pythia}. 
Also, this task simulates a common scenario in software development where a programmer starts writing a piece of code and relies on tooling to suggest how to continue. 
In addition, all CLMs support the code completion task.
In comparison, tasks such as code infill (\ie, infilling the missing code inside the code snippets) and \textit{code generation} (\ie, given textual descriptions, generating code) are not supported by all CLMs. Studying code completion tasks may achieve broader generalization.

Another reason for choosing code completion is that it could be conducted on the original collected code \textbf{\textit{without human intervention}}.
To minimize the impact of human factors on this study, it is prudent to preserve the authenticity of the collected code. Introducing artificial elements (\eg, natural language descriptions, code comments, and docstrings that are necessary for code generation) may \textbf{\textit{inadvertently influence the model's performance}} by providing additional context or clues that are not inherently part of the code's logic or structure. Therefore, we consistently automate the collection and preprocessing for all code groups, ensuring fair data processing and curation.

\subsection{\textbf{Code Complexity Metrics}}\label{sec:complexity}
To check the validity of the assumption of RQ1-RQ3, we consider two metrics to evaluate code complexity.

\subsubsection{\textbf{Cyclomatic complexity}}
Cyclomatic complexity~\cite{codeComplexity,codeMetrics} is a metric used to measure the complexity of a piece of software code. It quantifies the number of decision points or branches in a program, indicating the potential number of unique paths through the code. It is calculated based on the control flow graph of the code, which represents the flow of control between different {{statements, loops, conditionals, and function calls}}. It is determined by \textbf{\textit{counting the number of decision points in the graph}}, including conditional statements (if, switch), loops (for, while), and logical operators. We leverage Radon~\cite{radon} to calculate it on Python functions.

\subsubsection{\textbf{Cognitive complexity}}
Unlike cyclomatic complexity, which counts the number of decision points and branches in code, cognitive complexity~\cite{bieri1955cognitive} considers factors such as nesting levels, logical operators, and boolean expressions that can make code harder to understand.
It focuses on the structural complexity of the code and how it impacts human comprehension. In the implementation, we count the summation of the number of nesting levels, branches, and exceptions leveraging an off-the-shelf package~\cite{cog}.

\subsection{\textbf{Code Similarity Metrics}}\label{sec:simi}
Apart from code complexity metrics, we also consider six code similarity metrics to calculate the distance between code groups, further validating the assumption of RQ1-RQ3.

\subsubsection{\textbf{Levenshtein Similarity}}
Levenshtein distance measures the minimum number of operations required to transform one string into another. 
To project this distance into a similarity between 0 and 1, we normalize the distance by dividing the maximum string lengths of two strings.

\subsubsection{\textbf{CodeBleu Similarity}}
Code Bilingual Evaluation Understudy~\cite{codebleu} (CodeBLEU) measures the quality of machine-generated code against human reference code. 
Using n-grams of different lengths, CodeBLEU can assess the lexical structural and syntactic similarity between generated and reference codes.

\subsubsection{\textbf{Jaccard Similarity}}
Jaccard Similarity is a measure used to determine the similarity between two sets. It calculates the ratio of the size of the intersection of the sets to the size of their union. 
In the implementation, 
we follow the practice of existing works~\cite{neardup}, which use MinHash~\cite{minhash} to approximate the Jaccard Similarity between code snippets to boost efficiency.

\subsection{\textbf{Code Refactoring Operators}}
\label{sec:refactor-ops}
For RQ3, we consider five refactoring operators, covering \textit{\textbf{syntactic}} and \textbf{\textit{semantic}} refactoring.

\noindent \subsubsection{\textbf{Syntactic Refactoring Operators}} alters the code syntactic structure while the identifiers' names are untouched. 

\begin{itemize}
    \item \underline{If-condition flipping (IFF)}. We flip the condition of \texttt{if} statements and their true and false branches to syntactically restructure the program in our dataset while the original program logic and semantics are preserved.
    \item \underline{Loop transformation (Loop).} Similarly, we replace \texttt{while} loops with equivalent \texttt{for} loop structures, and vice versa, to syntactically refactor the code.
\end{itemize}

\subsubsection{\textbf{Semantic Refactoring Operators}} changes identifier names or adds additional context while the {code's functionality} remains unchanged. 

\begin{itemize}
    \item \underline{Identifier Renaming (Renm)}. To prevent CLMs from simply searching variable names, we leverage wordhoard~\cite{wordhoard} to fetch the synonyms of variable names and rename the variables in the Python code.
Note that we rename identifiers to synonyms instead of arbitrary names in that the variable names often indicate the code's functionality to be completed and serve as a necessary hint for CLMs to interpret the coding task and generate effective completions.
    \item \underline{Special Parameter Appending (Param).} We refactor functions by appending unnamed positional parameters (\texttt{*args}) and keyword parameters (\texttt{**kwargs}) in the parameter list of function declarations if they do not exist.
The appended parameters are unused in the function. Such refactoring does not change the behaviors of the function logic or any of its call sites while altering the function signature into a different form from what was seen in the CLMs' training set.
    \item {\underline{Performance Measurement Decoration (Deco).}} Since measuring the performance of a function, \eg, memory usage and execution time do not change the function's behavior, we also add performance measuring decorators to functions: \texttt{@timing}, which measures the total execution time of the decorated function, and \texttt{@measure\_memory\_usage}, which measures the memory usage of the function.
\end{itemize}

\subsection{\textbf{MIA (Membership Inference Attack)-related Metrics}}\label{sec:leakage-metrics}
For RQ4, we consider three existing metrics that may be used as indicators for data contamination, exploring whether they can distinguish different code groups. Note that these metrics are \textbf{\textit{model-dependent}}. Different models assign probabilities to the same input data, resulting in different scores.

\subsubsection{\textbf{Perplexity} (ppl)}
Perplexity~\cite{jelinek1977perplexity} is a measure used to evaluate the predictability of a sequence of words. Usually, the perplexity measures how ``surprised'' the model is to see a given value~\cite{carlini2019secret}. 
It is defined as the exponentiated average negative log-likelihood of a sequence. Given a tokenized sequence X with N tokens = ($x_0$, $x_1$, $\ldots$, $x_N$), the perplexity of $X$ is:
\begin{equation}
\begin{aligned}\label{eq:ppl}
    ppl(x) &= exp\{-\frac{1}{N}\sum^{N}_{i=1} log p_\theta (x_i|x_0\ldots x_{i-1})\}
\end{aligned}
\end{equation}
\noindent where $log~p_\theta$ is the log-likelihood of $x_i$ conditioned on the prefix $x_0\ldots x_{i-1})^{-1/N}$ according to the model. The lower the perplexity, the more natural the $X$ to the model. 

\subsubsection{\textbf{Perplexity$_{lower}$ (ppl\_lower)}}
A variation~\cite{topkmia2023} of the perplexity measure can be implemented by converting the input data to lowercase before calculating perplexity by removing the variability introduced by capitalization.

\subsubsection{\textbf{Zlib Compression Entropy}}
The Zlib of a given input $x$ denotes the number of bits after compressing $x$ with zlib. Zlib entropy~\cite{gailly2004zlib} of the text (\ie, the number of bits after compression with zlib) could be used as an indicator~\cite{extracting21,zhang2023ethicist}
\begin{equation}
\begin{aligned}\label{eq:zlib}
zlib_{en}(x) = {ppl(x)} / {len(zlib(x)}
\end{aligned}
\end{equation}

\subsubsection{\textbf{MIN-K\% PROB}}
\textsc{MIN-K\% PROB}~\cite{topkmia2023} was proposed to for membership inference attacks. It computes the score using the k\% of tokens with the lowest likelihoods.
\begin{equation}
\begin{aligned}\label{eq:min-k}
    MIN_{K\%}\ Prob(x) &= \frac{1}{E} \sum_{x_i\in Min_{K\%} (x)} log p(x_i|x_1, \ldots, x_{i-1})
\end{aligned}
\end{equation}
\noindent where E is the size of the Min$_{K\%(x)}$ set. In this study, we set K as $5.0$, $10.0$, and $20.0$ as set in previous work~\cite{topkmia2023}.

%% file: FigureTex/figure-rq123-design.tex
\begin{figure}[t]
    \centering
    \includegraphics[width=1.0\linewidth]{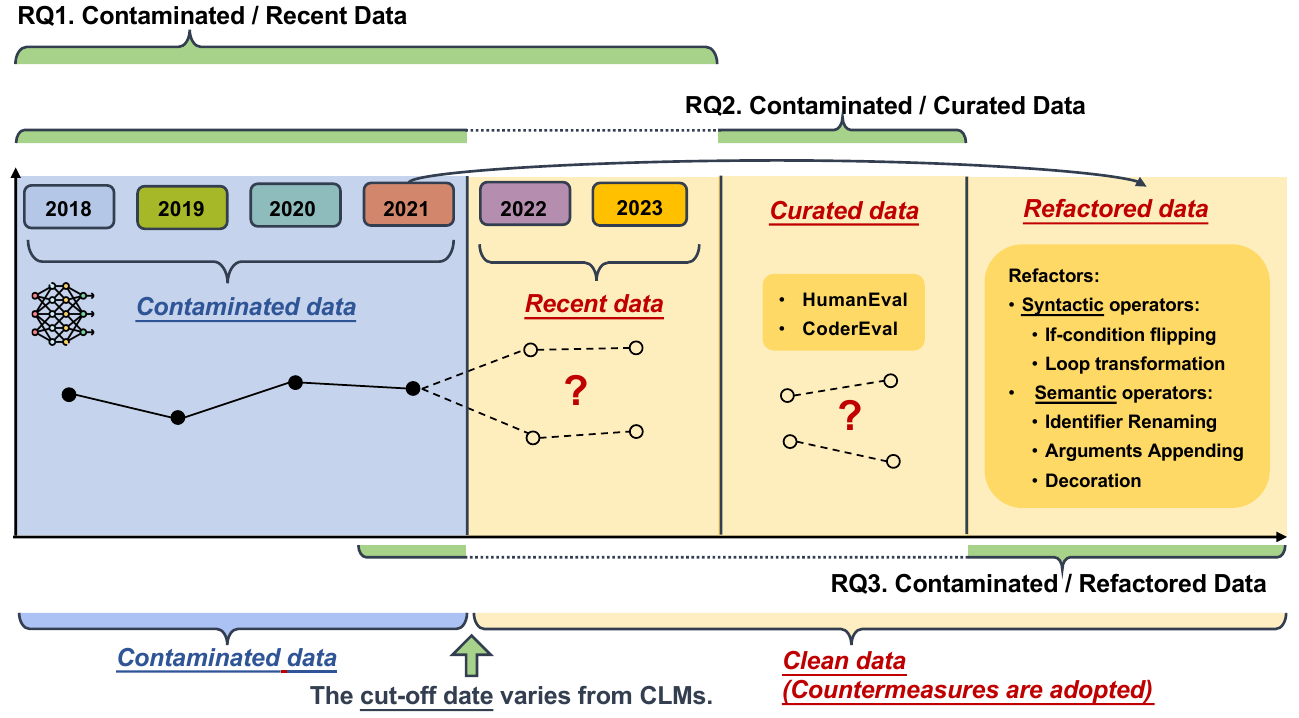}
    \setlength{\abovecaptionskip}{-0pt}
    \setlength{\belowcaptionskip}{-15pt}
    \caption{\textbf{Experiment Design for RQ1-3.} We assess CLMs' performance on contaminated and cleaned data. Three countermeasures (\ie, using recent data, using curated data or refactored contaminated data) are considered.}
    \label{fig:rq123-design}
\end{figure} 

%% file: FigureTex/figure-sim-design.tex
\begin{figure}[bt]
    \centering
    \includegraphics[width=1.0\linewidth]{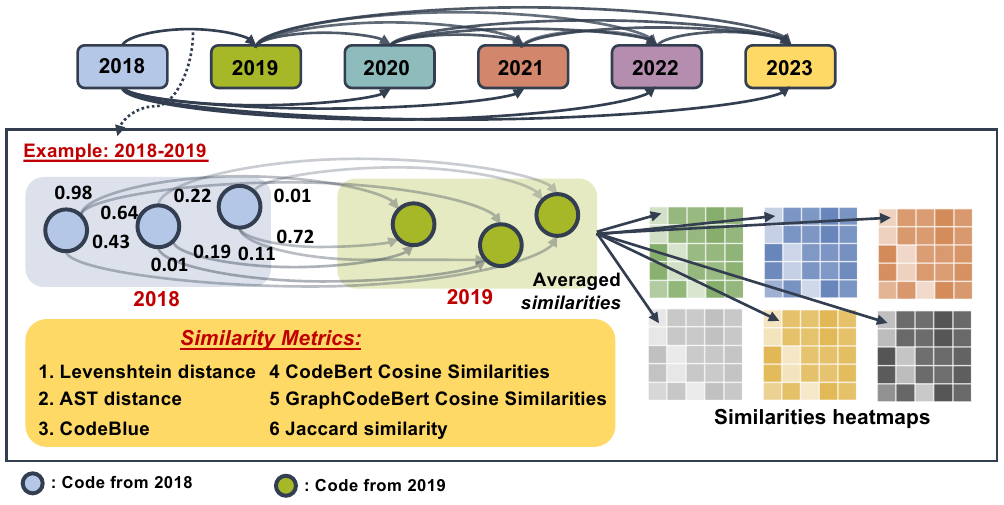}
    \setlength{\abovecaptionskip}{-10pt}
    \setlength{\belowcaptionskip}{-15pt}
    \caption{Pairwise Code Similarity Comparison}
    \label{fig:simi}
\end{figure} 

%% file: Tex/04-preperation.tex
\section{Experiment preparation}\label{sec:moti}

\subsection{Model and Training Data}\label{sec:model}
We select the state-of-the-art CLMs widely studied in recent code generation work. 
Table~\ref{tab:model} shows the \textit{\textbf{model information}} (including model name, base model, model size, and the first release time) and \textit{\textbf{training data information}} (including training source, the time span of the training data, and the languages the models support).
Note that since the majority of models do not specify the exact time of their training data, we then elaborate on how we determine the model's first release time, the data source, and the time span.

\input{FigureTex/figure_loc}

\input{Table/design}

On the one hand, we determined a model's \textbf{\textit{first release time}} according to the commit time when the model was uploaded. Note that since the first commit usually only updates the \texttt{README.md} without uploading the model, also, other files (\eg, licenses) could be uploaded before the model upload, so we carefully traced through the commit history in chronological order from past to present and located the first commit when the model checkpoints were uploaded. 
On the other hand, the \textbf{\textit{training source}} and time span of the training data are either determined by the model reports or inferred by their model release time (\ie, the cut-off date of the training data should at least goes before the model's first release). 

In the following, we list the detailed evidence of each model's model and data information.
\ding{202} \textbf{\textit{StarCoder}}~\cite{starcoder} explicitly states their training data, \textbf{\textit{the Stack}}~\cite{Kocetkov2022TheStack}. The model was first released on May 2023 according to the commit history~\cite{starcoderbase}. While \textit{the Stack} contains over 6TB of permissively licensed source code files covering 358 programming languages, 220.92M active GitHub repository names were collected from the event archives published between January 1st, 2015, and March 31st, 2022. 
\ding{203} The \textbf{\textit{StarChat}}~\cite{Tunstall2023starchat-alpha} is a fine-tuned version for assisting coding tasks. The model was released on June 7th, 2023~\cite{starchatbeta}. It is fine-tuned on OpenAssistant dataset~\cite{openassistant} whose upload date is April 12th, 2023 according to the commit time~\cite{oasst1}.
\ding{204} The \textbf{\textit{WizardCoder}}~\cite{wizardCoder23}~\cite{wizardcoder}, was fine-tuned based on StarCoder-15.5B. The training source includes the training data for \textit{StarCoder}, \ie, \textbf{\textit{the Stack}} and 20K instruction-following data used for fine-tuning the Code Alpaca model~\cite{codealpaca} whose upload date is May 13th, 2023~\cite{codealpacacommit}.
\ding{205} \textbf{\textit{CodeLlama-7b-Instruct}}~\cite{codellama7b}
was first uploaded on August 24th, 2023
~\cite{codellama7binstructcommit}. 
CodeLlama's training data is the same as that of Llama 2, which was trained between January 2023 and July 2023~\cite{llama27b}.
\ding{206} \textbf{\textit{Phind}}~\cite{phindcodellama34bv2} was fine-tuned on CodeLlama with an additional 1.5B tokens high-quality programming-related data. It was first released on Aug 29, 2023~\cite{phind} and since the cut-off-date of the additional data was disclosed, we assume this time as its training source cut-off-date. 
\ding{207} \textbf{ChatGLM}~\cite{du2022glm,zeng2022glm,chatglm} was first released on October 26th, 2023~\cite{chatglm36b}. Since the training source was not disclosed, we assume the training source was collected before October, 2023. 
\ding{208} \textbf{ChatGPT 3.5} has various variants. We use the \textit{ChatGPT3.5-turbo-0613} in the evaluation. Its first release time is June 2023, while its training data was cut off on May 2021~\cite{gpt35turbo}. 
\ding{209} \textbf{Github Copilot}~\cite{githubcopilot} was first released on June 29th, 2021~\cite{githubcopilotintroduce}. From November 30th, 2023, Copilot is empowered by GPT-4,~\cite{githubcopilotupdate} whose cut-off date is April 2023~\cite{gpt4turbo}.
\looseness=-1

\subsection{\textbf{Data Collection}}
RQ1 requires chronological data over a long period of time to construct annual code groups.
As such, we collect the Python code from January 1st, 2018, to December 31st, 2023.
Though there are several datasets available~\cite{Kocetkov2022TheStack}, they do not cover the most recent data (\eg, the latest six months). 
So, we first collect data from the~Stack~v2~\cite{thestackv2}, which contains code until March 31th, 2022 (Section~\ref{sec:collect-thestack}), and then collect newer data after March 31th, 2022, by crawling newly created repositories on GitHub (Section~\ref{sec:collect-repos}).

\subsubsection{\textbf{Data Collection from the~Stack~v2}}\label{sec:collect-thestack} 
We take the~Stack~v2~\cite{thestackv2} dataset as the code corpus for its representativeness and comprehensiveness. We use the deduplicated version of \texttt{the stack}, which contains over 2.9TB data of permissively licensed source code files covering 358 programming languages, from {January 1st, 2015} to {March 31st, 2022}. 
Typically, we take \texttt{Python} as our subject programming language for its popularity. 

\subsubsection{\textbf{Newer Data Collection}}
\label{sec:collect-repos}
To cover the code afterward (after March 31st, 2022), we crawl the Python repositories from GitHub with permissive licenses. 
In particular, we collect the repositories created from {April 1st, 2022} to {December 31st, 2023} with more than 50 stars. 
Note that we only collected repositories that are \textbf{created after} the time range of the~Stack~v2, so there should not be an overlap between the collected new data and the~Stack~v2.
As a result, a total of 15,743 repositories are collected. 

\subsubsection{\textbf{Data Processing}}
\label{sec:process}
We process the collected data by extracting functions from the Python code.
We filter out the functions that depend on other functions (\eg, the functions inside \texttt{classes} or inside other functions) or the empty functions (i.e.,~the function body is simply \texttt{pass}).

\subsubsection{\textbf{Statistics of Collected Data}}
After data collection and preparation, we collect a total of 12,493,174 Python functions in total.
The left subfigure of Fig.~\ref{fig:loc} shows the total number of functions collected yearly. 
The number of functions increases over years except for 2023.
The reason is that we only collect the data of 2023 from GitHub repositories with at least 50 stars, instead of all GitHub repositories like \textbf{\textit{the~Stack~v2}}.
As a result, the collected functions in 2023 are fewer than in previous years.
Nevertheless, since we conduct experiments on a sample of functions in each year and the total number of functions in 2023 is large enough (over 1 million), this difference in annual function numbers should not be a significant threat to our study.

Fig.~\ref{fig:loc} (right) shows the lines of code (LOCs) in 384 sampled functions each year. 
The lines of code in different years are essentially stable, with an average of 15.88 LOCs each year.

\subsection{\textbf{Annual Code Groups Preparation (RQ1)}}
Since there are over ten thousand code functions per year, traversing them all is unrealistic. 
Thus, we follow the practice of statistics of random sampling, setting a 95\% confidence level and 5\% margin of error (\ie, there is a 95\% confidence that the true population falls within $\pm$5\% of the sample estimate). 
It results in an average of 384 code functions annually. 
The sampled 384 functions in each year form the annual code groups in Fig.~\ref{fig:rq123-design}.
We refer to annual code groups as Code-2018, Code-2019, Code-2020, Code-2021, Code-2022, Code-2023, respectively.
\looseness=-1

\subsection{\textbf{Curated Code Groups Preparation (RQ2)}}
HumanEval~\cite{humaneval} dataset contains a set of Python functions, so we construct the Code-HumE code group directly with all functions in HumanEval.
CoderEval~\cite{yu2023codereval} dataset contains a set of Python files, so we extract functions from all files like in Section~\ref{sec:process} and then construct the Code-CodE code group.

\subsection{\textbf{Refactored Code Groups Preparation (RQ3)}}
The refactoring operators are expected to apply to contaminated code.
According to Table~\ref{tab:model}, the earliest cut-off date is September 2021, so we chose Code-2021 to apply the refactoring.
In particular, we apply the operators introduced in Section~\ref{sec:refactor-ops} to Code-2021. Each operator is applied to every function in Code-2021 where applicable. 
If an operator is not applicable (\eg., when attempting to flip if-else branches in code that does not contain an if-statement) to a certain code function under refactoring, we skip refactoring that function. 

If there are multiple ways to apply a particular operator, we randomly select one way to use it. For example, if one function has multiple if-conditions, we randomly pick one if-condition and apply the IFF (if-condition flipping) operator. 
The rationale behind employing a single operator at most once without repetition is that we want to analyze the impact of the most granular refactoring operator. Besides, by applying at most one type of operator to one function, we aim to isolate the influence of an individual operator rather than the compounded effects resulting from multiple operators.
We use Code-IFF, Code-Loop, Code-Renm, Code-Param, and Code-Deco to refer to the code group constructed by applying the five refactoring operations in Section~\ref{sec:refactor-ops}, respectively.

\subsection{\textbf{Coding Task Preparation}}\label{sec:task-prepare}
The code completion task understands the code prefix and generates the subsequent code. Since we want to minimize the human factors (\eg, manual annotations, subjective interpretations) in the code, we keep the code untouched without human curation and annotation. Indeed, without a natural language description, CLMs need to understand the code's intention from the code as-is and complete the suffix. 

To retain the code intent to the greatest extent, we only masked out the last statement in each function. Since we filtered out those functions with less than 3 lines when data preparation, this avoids the situation where the function body is empty after making out the last statement in the function. Furthermore, because only the final statement has been masked, the flexibility of code completion is relatively constrained. In other words, the correct ways to complete it should be limited, confined to completing only a single line of code. This setting also makes it easier to judge whether the models' output is correct. We use exact string matching to determine the outputs' correctness unless the last statement allows flexibility (such as a print statement). We ignore the concrete content for such cases and only match whether the print statement is predicted.

%% file: FigureTex/figure_loc.tex
\begin{figure}[t!]
    \centering
    \includegraphics[width=1.0\linewidth]{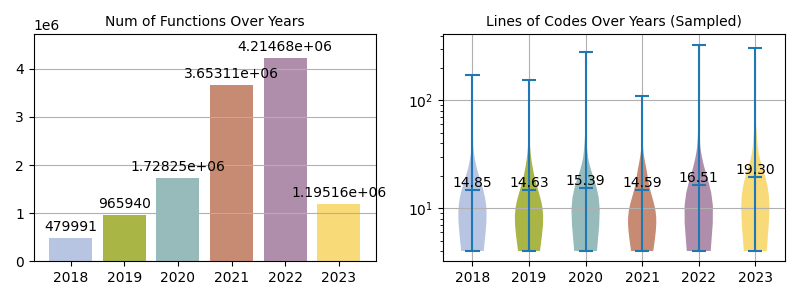}
    \setlength{\abovecaptionskip}{-10pt}
    \setlength{\belowcaptionskip}{-10pt}
    \caption{Statistics of Code Over Years}
    \label{fig:loc}
\end{figure}

%% file: Table/design.tex
\begin{table*}[t!]
    \centering
    \renewcommand\arraystretch{1.0}
    \caption{Large Language Models and Datasets}\label{tab:model}
    \resizebox{1\linewidth}{!}{
    \begin{tabular}{c|l|l|r|r|l|l|l}
    \toprule
    \multicolumn{8}{c}{\textbf{Code Large Language Models Statistics}} \\ 
    \midrule

    \multicolumn{5}{c}{\textbf{Model Information}} & \multicolumn{3}{|c}{\textbf{Data Information}} \\ \midrule
    \textbf{Index} & \textbf{Base Model} & \multicolumn{1}{c|}{\textbf{Model}} & \multicolumn{1}{c|}{\textbf{Size}} & \textbf{1st Release} & \multicolumn{1}{c}{\textbf{Source}} & \multicolumn{1}{|c}{\textbf{Time Span}}  & \multicolumn{1}{|c}{\textbf{Language}} \\ 
    \midrule
    
    {\large \ding{202}} & \multirow{-1}{*}{\textbf{StarCoder}} & \textbf{StarCoder} & 15.5B & May 2023 & the Stack~\cite{Kocetkov2022TheStack} & Jan 2015$\sim$ Mar 2022  & 358 Programming languages \\
    
    {\large \ding{203}} &  & \textbf{StarChat} & 15.5B & Jun 2023 & GitHub & Up to April 2023 & \begin{tabular}[l]{@{}l@{}}80+ programming languages.\end{tabular} \\ 
    
    {\large \ding{204}} & & \textbf{WizardCoder} & 15.5B & Jun 2023 & the Stack~\cite{Kocetkov2022TheStack}, CodeAlpaca-20k & Jan 2015$\sim$ Mar 2023  & 358 Programming languages and English \\
    \hline
    
    {\large \ding{205}} & & \textbf{CodeLlama-Instruct} & 7B  & Apr 2023 & GitHub + StackOverflow & Up to Jan 2023 & \begin{tabular}[l]{@{}c@{}}Python, C/C++, TypeScript, Java and more\end{tabular} \\
     
    {\large \ding{206}} & \multirow{-2}{*}{\textbf{Llama2}} & \textbf{Phind-CodeLlama-34b} & 34B & Aug 2023 & GitHub + StackOverflow & Up to Aug 2023 & \begin{tabular}[l]{@{}c@{}}Python, C/C++, TypeScript, Java and more\end{tabular} \\
    \hline
    
    

    \hline
    {\large \ding{207}} & \textbf{GLM} & \textbf{ChatGLM2} & 6B & Oct 2023 & -- & Up to May 2022  & Python, Java, JavaScript \\
    
    \hline
    {\large \ding{208}} & & \textbf{ChatGPT 3.5-turbo} & -- & June 2023 & Public Data & Up to Sep 2021  & 95+ Programming languages \\
     
    {\large \ding{209}} & \multirow{-2}{*}{\textbf{ChatGPT}} & \textbf{Github-Copilot} & -- & Jun 2021 & Public Data & Up to April 2023  & 95+ Programming languages \\
\bottomrule
    \end{tabular}
    }
\end{table*}

%% file: Tex/05-evaluation-rq1.tex
\section{Evaluation}\label{sec:methd}

\input{FigureTex/figure_refactor_complexity}

We use nucleus sampling~\cite{nuclear} in line with recent works~\cite{du2023classeval,yu2023codereval,ouyang2023llm}, where five solution samples are randomly generated with a temperature of 0.2~\cite{humaneval}.
The experiments are conducted on a computational infrastructure comprising two NVIDIA RTX 6000 Ada GPUs, each with 48GB of graphics memory. 

\input{FigureTex/figure-similarity-heatmap}

\subsection{\textbf{Evaluation on the Assumption of RQ1-RQ3}}\label{sec:assumption}
Before we start RQ1-RQ3, which uses CLMs' performance as an indicator for data contamination, there is an assumption that the contaminated/cleansed data are on similar difficulty levels. 
Thus, we demonstrate the code complexity of code groups and show the similarities between code groups.

The code complexities over code groups are shown in Fig.~\ref{fig:refactor-complex}. 
The average cyclomatic and cognitive complexities over code groups are generally similar, with slight fluctuation. 
In particular, among Code-2018 to Code-2023 (groups before the first dashed vertical line), \textbf{\textit{a slight ascending trend}} could be observed on both complexities, reaching the highest mean and medium complexities of Code-2023. 
For benchmarks (between the first and second vertical dashed lines), the complexities of Code-CodE are higher than that of Code-HumE in both metrics, meaning that the codes in Code-CodE are slightly more complex than Code-HumE on average, which is in line with the observation in previous work~\cite{yu2023codereval}.

Among the refactored code groups (after the second vertical line), the Code-IFF and Code-Loop have the highest scores on both complexities, while other factored groups are similar to Code-2018 to Code-2021 with small fluctuations.

\begin{mdframed}[style=MyFrame]
\textit{\textbf{Finding 1:}}
The code groups display \textbf{\textit{similar code complexity}} in terms of cyclomatic and cognitive complexities, validating the assumptions for RQ1-RQ3 despite minor fluctuations. 
\\
\textit{\textbf{Finding 2:}}
Code complexity across different years appears to be relatively stable, exhibiting a slight upward trend.
\\
\textit{\textbf{Finding 3:}}
The code complexity of CoderEval is slightly higher than early code groups, while HumanEval is at a similar complexity level as early codes.
\end{mdframed}

We further measure the three code similarities over various code groups. The results of Levenshtein and Jaccard distance are visualized in Fig.~\ref{fig:rq-sim}. The CodeBleu stays at an extremely low level (from 0.000018 to 0.009587), so we omit the visualization of it.
The heatmaps show that the similarities over code groups do not differ substantially. 
Also, clear that the similarities between code groups are relatively low, ranging from 0.18 to 0.20 (Levenshtein similarity) and 0.05 to 0.06 (Jaccard similarity). Note that the existing work~\cite{thestackv2} uses a threshold of 0.85 Jaccard similarity. The average results over these code groups are significantly lower, indicating the overlapping degree among these code groups are at a relatively minimal level.
Similarity metrics among the code groups do not differ substantially. 
\looseness=-1

\input{Table/rq1}

\begin{mdframed}[style=MyFrame]
\textit{\textbf{Finding 4:}}
The various code groups exhibit a \textbf{\textit{consistent level of similarity}} among themselves, which echoes the observation on code complexity.
\end{mdframed}

\vspace{-16pt}
\subsection{RQ1. How does CLMs' performance differ on contaminated/recent data?}

Table~\ref{tab:rq1} (from entries ``Code-2018'' to ``Code-2023-oct'') shows Pass@k with nucleus sampling on code groups from different years. Overall, Copilot consistently outperforms other models in all the code groups, followed by Phind. 
Horizontally, the value of Pass@k \textit{stays stable from Code-2018 to Code-2022}, followed by \textbf{\textit{a slight decrease in Code-2022}} observed in all CLMs. Combining the complexity analysis in Fig.~\ref{fig:refactor-complex}, it is reasonable because Code-2022 has higher code complexities than previous code groups. 

An interesting situation happens in Code-2023. Though most CLMs were released before 2023 (meaning that these CLMs \textbf{\textit{are less likely}} to see these codes), and the complexity of Code-2023 is higher than the earlier code groups, the \textbf{\textit{Pass@k in Code-2023 consistently outperforms earlier code groups over all the CLMs.}} 
This observation is counter-intuitive because the general belief is that the more recent data suffer less from data contamination. 
So, we further sampled a code group after Oct 2023, which is later than \textbf{all} CLMs' cut-off date, to check whether the observation still holds. See the column ``Code-2023-oct''; the results \textbf{\textit{echo}} that in Code-2023 again. CLMs even achieve higher performance in Code-2023-oct. 

A possible reason behind this is the popularity of AI programming assistants such as Copilot. As reported by Github in June 2022~\cite{githubcopilotavailable}, nearly 40\% of the code is being written by GitHub Copilot in popular coding languages like Python in last one year. That means the code created and committed in 2023 may be initially predicted by AI; it stands to reason that CLMs perform better on them.

\begin{mdframed}[style=MyFrame]
\textit{\textbf{Finding 5:}} CLMs do not necessarily perform worse on recent data (\ie, later than their release). On the contrary, the CLMs may perform better on recent codes. 
It indicates that \textbf{\textit{using more recent data may not be an ideal strategy}} to alleviate data contamination. 
\\
\textit{\textbf{Implication:}} The popularity of AI programming assistants such as Copilot may further \textbf{\textit{exacerbate data contamination threats}}. It also indicates that simply amassing more recent data may not be optimal when creating new benchmarks.
\end{mdframed}

%% file: FigureTex/figure_refactor_complexity.tex
\begin{figure}[t!]
    \centering
    \includegraphics[width=0.98\linewidth]{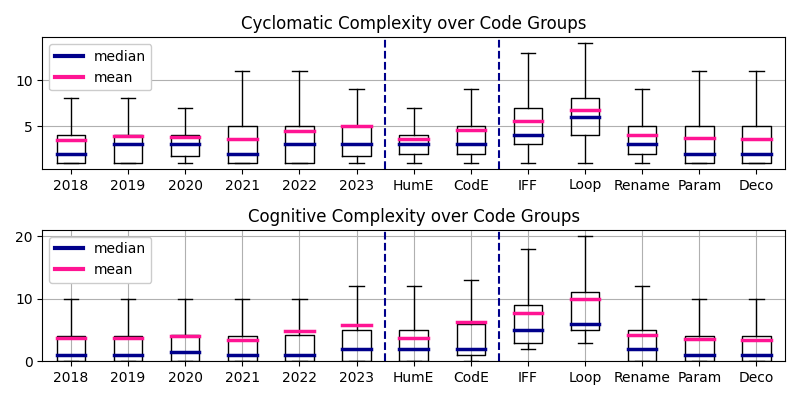}
    \setlength{\abovecaptionskip}{-10pt}
    \setlength{\belowcaptionskip}{-10pt}
    \caption{Complexity Comparison of Various Code Groups
    }
    \label{fig:refactor-complex}
\end{figure} 

%% file: FigureTex/figure-similarity-heatmap.tex
\begin{figure}[t!]
    \centering
    \includegraphics[width=0.98\linewidth]{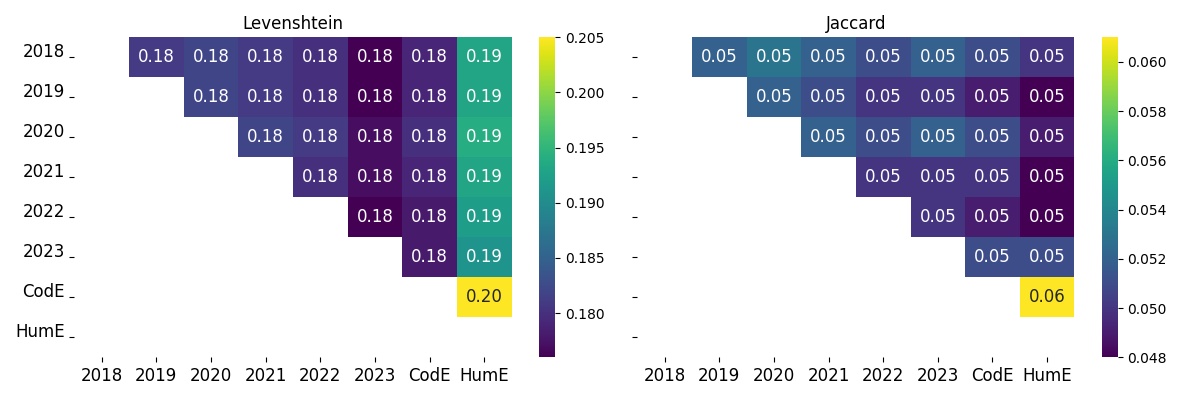}
    \setlength{\abovecaptionskip}{-5pt}
    \setlength{\belowcaptionskip}{-10pt}
    \caption{Code Similarities Over Various Code Groups}
    \label{fig:rq-sim}
\end{figure}

%% file: Table/rq1.tex
\begin{table*}[ht]
\centering
\renewcommand\arraystretch{1.21}
\caption{\textbf{Pass@k with Nucleus Sampling on Contaminated/Recent/Curated Code Groups.}
The \colorbox[HTML]{97C3C2}{greener}, the larger. 
We \textbf{\underline{underline}} the results on the data collected \textit{after}
each model's release time (\ie, \textbf{\textit{cleansed}}) to facilitate clearer demonstration.
}
\label{tab:rq1}
\resizebox{1\textwidth}{!}{
\begin{tabular}{lrrr|rrr|rrr|rrr|rrr|rrr||rrr||rrr|rrr}
\toprule
 & \multicolumn{3}{c}{\textbf{Code-2018}} & \multicolumn{3}{c}{\textbf{Code-2019}} & \multicolumn{3}{c}{\textbf{Code-2020}} & \multicolumn{3}{c}{\textbf{Code-2021}} & \multicolumn{3}{c}{\textbf{Code-2022}} & \multicolumn{3}{c||}{\textbf{Code-2023}} & \multicolumn{3}{c||}{\textbf{Code-2023-oct}} & \multicolumn{3}{c}{\textbf{Code-CodE}} & \multicolumn{3}{c}{\textbf{Code-HumE}} \\
\hline
\textbf{Model} & P@1 & P@3 & P@5 & P@1 & P@3 & P@5 & P@1 & P@3 & P@5 & P@1 & P@3 & P@5 & P@1 & P@3 & P@5 & P@1 & P@3 & P@5 & P@1 & P@3 & P@5 & P@1 & P@3 & P@5 & P@1 & P@3 & P@5 \\
 \midrule
\textbf{StarCoder} & \cellcolor[HTML]{E7F1F1}30.70 & \cellcolor[HTML]{E2EEEE}32.60 & \cellcolor[HTML]{E0EDED}33.60 & \cellcolor[HTML]{E8F2F1}30.50 & \cellcolor[HTML]{E4EFEF}32.00 & \cellcolor[HTML]{E2EEEE}32.80 & \cellcolor[HTML]{E9F2F2}30.20 & \cellcolor[HTML]{E6F1F0}31.20 & \cellcolor[HTML]{E1EEED}33.10 & \cellcolor[HTML]{EBF3F3}29.40 & \cellcolor[HTML]{E9F2F2}30.20 & \cellcolor[HTML]{E6F1F0}31.20 & \cellcolor[HTML]{F0F7F6}27.30 & \cellcolor[HTML]{EBF3F3}29.40 & \cellcolor[HTML]{E8F2F1}30.50 & \cellcolor[HTML]{DEECEC}\underline{34.10} & \cellcolor[HTML]{DBEAEA}\underline{35.40} & \cellcolor[HTML]{D8E9E8}\underline{36.50} & \cellcolor[HTML]{DFECEC}\underline{33.90} & \cellcolor[HTML]{D8E8E8}\underline{36.70} & \cellcolor[HTML]{D5E7E6}\underline{37.80} & \cellcolor[HTML]{CDE2E1}\underline{40.90} & \cellcolor[HTML]{CBE1E0}\underline{41.70} & \cellcolor[HTML]{C8DFDF}\underline{42.60} & \cellcolor[HTML]{B2D2D1}51.20 & \cellcolor[HTML]{AACDCD}54.30 & \cellcolor[HTML]{A8CDCC}54.90 \\
\textbf{StarChat} & \cellcolor[HTML]{D6E7E7}37.20 & \cellcolor[HTML]{D0E4E3}39.80 & \cellcolor[HTML]{CEE2E2}40.60 & \cellcolor[HTML]{DEECEB}34.40 & \cellcolor[HTML]{D7E8E7}37.00 & \cellcolor[HTML]{D4E6E6}38.00 & \cellcolor[HTML]{DAEAE9}35.70 & \cellcolor[HTML]{D2E5E5}38.80 & \cellcolor[HTML]{D0E4E3}39.60 & \cellcolor[HTML]{D3E5E5}38.50 & \cellcolor[HTML]{CCE2E1}41.10 & \cellcolor[HTML]{C9E0DF}42.20 & \cellcolor[HTML]{E2EEEE}32.60 & \cellcolor[HTML]{DBEAEA}35.40 & \cellcolor[HTML]{D6E7E7}37.20 & \cellcolor[HTML]{D1E4E4}39.30 & \cellcolor[HTML]{CCE2E1}41.10 & \cellcolor[HTML]{CBE1E0}41.70 & \cellcolor[HTML]{CAE0E0}\underline{41.90} & \cellcolor[HTML]{CAE0E0}\underline{41.90} & \cellcolor[HTML]{CAE0E0}\underline{41.90} & \cellcolor[HTML]{B8D6D5}48.70 & \cellcolor[HTML]{AED0CF}52.60 & \cellcolor[HTML]{ACCFCE}53.50 & \cellcolor[HTML]{B8D6D5}48.80 & \cellcolor[HTML]{A8CDCC}54.90 & \cellcolor[HTML]{A8CDCC}54.90 \\
\textbf{WizardCoder} & \cellcolor[HTML]{D3E5E5}38.50 & \cellcolor[HTML]{D0E4E3}39.60 & \cellcolor[HTML]{CDE2E1}40.90 & \cellcolor[HTML]{DCEBEB}34.90 & \cellcolor[HTML]{D6E7E7}37.50 & \cellcolor[HTML]{D5E7E6}37.80 & \cellcolor[HTML]{D3E5E5}38.50 & \cellcolor[HTML]{CEE3E2}40.40 & \cellcolor[HTML]{CDE2E1}40.90 & \cellcolor[HTML]{CEE3E2}40.40 & \cellcolor[HTML]{C7DEDE}43.20 & \cellcolor[HTML]{C5DDDD}43.80 & \cellcolor[HTML]{DEECEC}34.10 & \cellcolor[HTML]{D9E9E9}36.20 & \cellcolor[HTML]{D6E7E7}37.20 & \cellcolor[HTML]{CEE2E2}40.60 & \cellcolor[HTML]{CAE0E0}41.90 & \cellcolor[HTML]{C9E0DF}42.40 & \cellcolor[HTML]{C5DDDD}\underline{43.80} & \cellcolor[HTML]{BCD8D7}\underline{47.40} & \cellcolor[HTML]{B9D6D6}\underline{48.40} & \cellcolor[HTML]{AFD1D0}52.20 & \cellcolor[HTML]{A4CAC9}56.50 & \cellcolor[HTML]{9FC7C7}58.30 & \cellcolor[HTML]{A7CCCB}55.50 & \cellcolor[HTML]{A2C9C8}57.30 & \cellcolor[HTML]{A0C8C7}57.90 \\
\textbf{Codellama} & \cellcolor[HTML]{D7E8E7}37.00 & \cellcolor[HTML]{D0E4E3}39.60 & \cellcolor[HTML]{CEE2E2}40.60 & \cellcolor[HTML]{D6E7E7}37.20 & \cellcolor[HTML]{D0E4E3}39.80 & \cellcolor[HTML]{CFE3E3}40.10 & \cellcolor[HTML]{D7E8E7}37.00 & \cellcolor[HTML]{CEE2E2}40.60 & \cellcolor[HTML]{CCE2E1}41.10 & \cellcolor[HTML]{D3E5E5}38.50 & \cellcolor[HTML]{CEE2E2}40.60 & \cellcolor[HTML]{CCE2E1}41.10 & \cellcolor[HTML]{E0EDED}33.60 & \cellcolor[HTML]{DAE9E9}35.90 & \cellcolor[HTML]{D7E8E7}37.00 & \cellcolor[HTML]{D4E6E6}\underline{38.00} & \cellcolor[HTML]{CEE3E2}\underline{40.40} & \cellcolor[HTML]{CBE1E1}\underline{41.40} & \cellcolor[HTML]{C7DEDE}\underline{43.20} & \cellcolor[HTML]{BED9D9}\underline{46.40} & \cellcolor[HTML]{BED9D9}\underline{46.60} & \cellcolor[HTML]{AFD1D0}{52.20} & \cellcolor[HTML]{A6CBCB}{55.70} & \cellcolor[HTML]{A3C9C9}{57.00} & \cellcolor[HTML]{ADCFCF}{53.00} & \cellcolor[HTML]{AACDCD}{54.30} & \cellcolor[HTML]{A8CDCC}{54.90} \\
\textbf{Phind} & \cellcolor[HTML]{CBE1E1}41.40 & \cellcolor[HTML]{C8DFDF}42.70 & \cellcolor[HTML]{C5DDDD}43.80 & \cellcolor[HTML]{CDE2E1}40.90 & \cellcolor[HTML]{C6DEDD}43.50 & \cellcolor[HTML]{C5DDDD}44.00 & \cellcolor[HTML]{C5DDDD}44.00 & \cellcolor[HTML]{C0DADA}45.80 & \cellcolor[HTML]{BED9D9}46.40 & \cellcolor[HTML]{C4DDDC}44.30 & \cellcolor[HTML]{BDD9D8}46.90 & \cellcolor[HTML]{BCD8D7}47.40 & \cellcolor[HTML]{D2E5E5}38.80 & \cellcolor[HTML]{CBE1E1}41.40 & \cellcolor[HTML]{C9E0DF}42.40 & \cellcolor[HTML]{D2E5E5}38.80 & \cellcolor[HTML]{D0E4E3}39.60 & \cellcolor[HTML]{D0E4E3}39.60 & \cellcolor[HTML]{BED9D9}\underline{46.40} & \cellcolor[HTML]{BCD8D7}\underline{47.40} & \cellcolor[HTML]{BCD8D7}\underline{47.40} & \cellcolor[HTML]{99C3C3}60.90 & \cellcolor[HTML]{99C3C3}60.90 & \cellcolor[HTML]{98C3C2}61.30 & \cellcolor[HTML]{A5CBCA}56.10 & \cellcolor[HTML]{A2C9C8}57.30 & \cellcolor[HTML]{A0C8C7}57.90 \\
\textbf{ChatGLM2} & \cellcolor[HTML]{FCFEFD}22.70 & \cellcolor[HTML]{EDF5F4}28.60 & \cellcolor[HTML]{E2EEEE}32.60 & \cellcolor[HTML]{F8FBFB}24.20 & \cellcolor[HTML]{F4F9F8}26.00 & \cellcolor[HTML]{F0F7F6}27.30 & \cellcolor[HTML]{F8FBFB}24.20 & \cellcolor[HTML]{F1F7F7}26.80 & \cellcolor[HTML]{EFF6F6}27.60 & \cellcolor[HTML]{FFFFFF}21.40 & \cellcolor[HTML]{F7FAFA}24.70 & \cellcolor[HTML]{F5F9F9}25.50 & \cellcolor[HTML]{FEFEFE}22.10 & \cellcolor[HTML]{F9FCFB}24.00 & \cellcolor[HTML]{F6FAFA}25.00 & \cellcolor[HTML]{FCFEFD}\underline{22.70} & \cellcolor[HTML]{F5FAF9}\underline{25.30} & \cellcolor[HTML]{F1F7F7}\underline{26.80} & \cellcolor[HTML]{F0F7F6}\underline{27.30} & \cellcolor[HTML]{EEF5F5}\underline{28.10} & \cellcolor[HTML]{E8F2F1}\underline{30.50} & \cellcolor[HTML]{E7F1F1}\underline{30.90} & \cellcolor[HTML]{E2EEEE}\underline{32.60} & \cellcolor[HTML]{DDEBEB}\underline{34.80} & \cellcolor[HTML]{DDEBEB}34.80 & \cellcolor[HTML]{D9E9E9}36.00 & \cellcolor[HTML]{D6E7E7}37.20 \\
\textbf{ChatGPT-3.5} & \cellcolor[HTML]{E3EFEF}32.30 & \cellcolor[HTML]{DEECEC}34.10 & \cellcolor[HTML]{DEECEB}34.40 & \cellcolor[HTML]{E9F2F2}30.20 & \cellcolor[HTML]{E4F0EF}31.80 & \cellcolor[HTML]{E2EEEE}32.60 & \cellcolor[HTML]{E7F1F1}30.70 & \cellcolor[HTML]{E7F1F1}31.00 & \cellcolor[HTML]{E6F1F0}31.20 & \cellcolor[HTML]{E4F0EF}31.80 & \cellcolor[HTML]{E1EDED}33.30 & \cellcolor[HTML]{DFECEC}33.90 & \cellcolor[HTML]{EEF5F5}\underline{28.10} & \cellcolor[HTML]{E9F2F2}\underline{30.20} & \cellcolor[HTML]{E9F2F2}\underline{30.20} & \cellcolor[HTML]{DEECEC}\underline{34.10} & \cellcolor[HTML]{DCEBEA}\underline{35.20} & \cellcolor[HTML]{DBEAEA}\underline{35.40} & \cellcolor[HTML]{D6E7E7}\underline{37.50} & \cellcolor[HTML]{D1E5E4}\underline{39.10} & \cellcolor[HTML]{D0E4E3}\underline{39.80} & \cellcolor[HTML]{BCD8D7}\underline{47.40} & \cellcolor[HTML]{B8D6D5}\underline{48.70} & \cellcolor[HTML]{B7D5D5}\underline{49.10} & \cellcolor[HTML]{B0D1D1}51.80 & \cellcolor[HTML]{AACDCD}54.30 & \cellcolor[HTML]{A7CCCB}55.50 \\
\textbf{Github-Copilot} & \cellcolor[HTML]{B9D6D6}48.40 & \cellcolor[HTML]{B6D4D4}49.70 & \cellcolor[HTML]{B4D3D3}50.50 & \cellcolor[HTML]{B8D6D5}49.00 & \cellcolor[HTML]{B6D4D4}49.70 & \cellcolor[HTML]{B4D4D3}50.30 & \cellcolor[HTML]{B8D6D5}48.70 & \cellcolor[HTML]{B3D3D2}50.80 & \cellcolor[HTML]{B1D2D1}51.60 & \cellcolor[HTML]{ACCFCE}53.40 & \cellcolor[HTML]{A7CCCB}55.50 & \cellcolor[HTML]{A5CBCA}56.20 & \cellcolor[HTML]{C3DCDB}44.80 & \cellcolor[HTML]{BFDAD9}46.10 & \cellcolor[HTML]{BED9D9}46.60 & \cellcolor[HTML]{B4D3D3}50.50 & \cellcolor[HTML]{AED0CF}52.60 & \cellcolor[HTML]{ADCFCF}53.10 & \cellcolor[HTML]{A5CBCA}\underline{56.00} & \cellcolor[HTML]{9FC7C7}\underline{58.30} & \cellcolor[HTML]{9EC6C6}\underline{58.90} & \cellcolor[HTML]{9BC5C4}60.00 & \cellcolor[HTML]{93C0BF}63.00 & \cellcolor[HTML]{8EBDBC}64.80 & \cellcolor[HTML]{A0C8C7}57.90 & \cellcolor[HTML]{97C2C1}61.60 & \cellcolor[HTML]{95C1C1}62.20

 \\
\bottomrule
\end{tabular}
}
\end{table*}

%% file: Tex/05-evaluation-rq2.tex
\subsection{RQ2. How does CLMs' performance differ on contaminated/curated data?}\label{sec:rq2}

The performance on the curated datasets (\ie, HumanEval~\cite{humaneval} and CoderEval~\cite{yu2023codereval}) is shown in the last two entries (\ie, ``Code-CodE'' and ``Code-HumE'') in Table~\ref{tab:rq1}. 
\textbf{\textit{
In general, CLMs achieve better performance on curated datasets compared to that on the contaminated data in terms of Pass@k scores. 
}}
Interestingly, the performance of CLMs is significantly better on CoderEval even though its release time is later than most models' release time. 
This may be because these curated benchmarks are carefully screened and processed manually, the code quality is relatively high, and it is more likely to conform to the distribution of instruction data of CLMs that has been supervised and fine-tuned.
Such results indicate that the program comprehension and completion ability of CLMs are effectively generalized to datasets curated after the models' cut-off date. 

\begin{mdframed}[style=MyFrame]
\textit{\textbf{Finding 6:}} 
CLMs perform better on recently curated datasets, compared to the performance on the contaminated code groups.
It indicates that \textbf{\textit{using curated datasets may not be effective}} in mitigating data contamination threats.
\end{mdframed}

%% file: Tex/05-evaluation-rq3.tex
\subsection{RQ3. How does CLMs' performance differ on contaminated/refactored data?}

\input{Table/rq3}

Table~\ref{tab:rq3} shows the Pass@1 scores of different CLMs on the code group Code-2021 and various refactored code groups.

\noindent \textbf{\textit{Syntactic Refactoring Operators.}}
The change in the Pass@1 of CLMs varies.
\texttt{IFF} is capable of degrading the Pass@1 score of some CLMs while loop transformation may not be effective. 
The possible reason behind this may be that such syntactic refactoring may or may not make the code close to the data distribution of the training set. On the contrary, syntactic refactoring could even upgrade the model's performance (Code-Loop). 
Hence, the performance may vary.
\looseness=-1

\noindent \textbf{\textit{Semantic Refactoring Operators.}}
The Pass@1 score of CLMs decreases after refactoring.
Renaming identifiers with their synonyms decreases the Pass@1 scores of CLMs in that the new identifier name, although it is the synonym of the original name, may convey different contextual meanings in the code. 
For instance, in the following code snippet \texttt{encoded = base64.b64encode(io.open(mp4, 'r+b').read())}, a video file is read and stored as base64 encoding in variable \texttt{encoded}.
The refactoring operator renames the variable to \texttt{ciphered}.
Although \texttt{ciphered} is a synonym of \texttt{encoded} in natural language, it does not fit in the coding context since there is no cryptographic ciphering procedure. 
As a result, the renaming may alter the semantics of the program and provide different hints to CLMs to complete the code, so that the CMLs' performance is influenced.

Special parameter appending and performance measurement decoration prepend additional contexts to CLMs' prompt but the original functionalities of the program are left unchanged. 
We observe that such refactoring operators are effective in degrading the CLMs' performance.

\begin{mdframed}[style=MyFrame]
\textit{\textbf{Finding 7:}} 
\textbf{\textit{Syntactic refactoring}} may not be useful to alleviate data contamination. On the contrary, refactoring the code structure could \textbf{\textit{even upgrade}} the models' performance. 
\\
\textit{\textbf{Implication:}} 
Semantic code refactoring operators (\texttt{Renm}, \texttt{Param} and \texttt{Deco}) could be more effective in evaluating the data contamination threats of CLMs.
\end{mdframed}

%% file: Table/rq3.tex
\begin{table}[t]
\centering
\renewcommand\arraystretch{1.1}
\caption{\textbf{Pass@1 Results (Greedy) on Original (Code-2021) and Various Refactored Code Groups.} 
The entries $\Delta$ is the difference between the previous column and the ``Origin'' column. 
Foreground color \down{blue} means decrease and \upp{red} means increase. The background color highlights the value of the Pass@1 result: the  \colorbox[HTML]{97C3C2}{greener}, the larger. }
\label{tab:rq3}
\resizebox{1\linewidth}{!}{
\begin{tabular}{l|r|rrrrrrrrrr}
\toprule
 & \multicolumn{1}{c|}{\textbf{2021}} & \multicolumn{2}{c}{\textbf{Code-IFF}} & \multicolumn{2}{c}{\textbf{Code-Loop}} & \multicolumn{2}{c}{\textbf{Code-Renm}} & \multicolumn{2}{c}{\textbf{Code-Param}} & \multicolumn{2}{c}{\textbf{Code-Deco}} \\ \cline{2-12} 
Model & \multicolumn{1}{c|}{{P@1}} & \multicolumn{1}{c}{{ {P@1}}} & \multicolumn{1}{c}{$\Delta$} & \multicolumn{1}{c}{{ {P@1}}} & \multicolumn{1}{c}{$\Delta$} & \multicolumn{1}{c}{{ {P@1}}} & \multicolumn{1}{c}{$\Delta$} & \multicolumn{1}{c}{{ {P@1}}} & \multicolumn{1}{c}{$\Delta$} & \multicolumn{1}{c}{{ {P@1}}} & \multicolumn{1}{c}{$\Delta$} \\ \hline

Codellama & \cellcolor[HTML]{D0E4E3}38.5 & \cellcolor[HTML]{D9E9E8}35.1 & \down{-3.4} & \cellcolor[HTML]{C8DFDF}41.8 & \upp{3.3} & \cellcolor[HTML]{D1E4E4}38.2 & \down{-0.3} & \cellcolor[HTML]{D3E6E5}37.3 & \down{-1.2} & \cellcolor[HTML]{F6FAFA}23.4 & \down{-15}.1 \\
StarCode & \cellcolor[HTML]{E7F1F1}29.4 & \cellcolor[HTML]{E4EFEF}30.5 & \upp{1.1} & \cellcolor[HTML]{DAEAE9}34.4 & \upp{5.0} & \cellcolor[HTML]{E4EFEF}30.5 & \upp{1.1} & \cellcolor[HTML]{EBF4F4}27.6 & \down{-1.8} & \cellcolor[HTML]{F3F8F8}24.7 & \down{-4.7} \\
StarCoder-chat & \cellcolor[HTML]{D0E4E3}38.5 & \cellcolor[HTML]{D0E4E3}38.5 & \equ{0.0} & \cellcolor[HTML]{CCE1E1}40.2 & \upp{1.7} & \cellcolor[HTML]{D4E6E5}37.1 & \down{-1.4} & \cellcolor[HTML]{D5E6E6}36.7 & \down{-1.8} & \cellcolor[HTML]{E4EFEF}30.5 & \down{-8.0} \\
Wizardcode & \cellcolor[HTML]{CBE1E1}40.4 & \cellcolor[HTML]{CFE3E3}39.1 & \down{-1.3} & \cellcolor[HTML]{BAD7D6}47.5 & \upp{7.1} & \cellcolor[HTML]{CBE1E1}40.4 & \equ{0.0} & \cellcolor[HTML]{CCE1E1}40.2 & \down{-0.2} & \cellcolor[HTML]{F3F8F8}24.5 & \down{-15.9} \\
Phind & \cellcolor[HTML]{C2DBDB}44.3 & \cellcolor[HTML]{B8D6D5}48.3 & \upp{4.0} & \cellcolor[HTML]{ADCFCF}52.5 & \upp{8.2} & \cellcolor[HTML]{C4DDDC}43.3 & \down{-1.0} & \cellcolor[HTML]{C6DEDD}42.5 & \down{-1.8} & \cellcolor[HTML]{CFE3E3}39.1 & \down{-5.2} \\
Chatglm2 & \cellcolor[HTML]{FBFDFD}21.4 & \cellcolor[HTML]{FDFEFE}20.7 & \down{-0.7} & \cellcolor[HTML]{F5F9F9}23.8 & \upp{2.4} & \cellcolor[HTML]{FAFCFC}21.8 & \upp{0.4} & \cellcolor[HTML]{FEFEFE}20.2 & \down{-1.2} & \cellcolor[HTML]{FFFFFF}19.5 & \down{-1.9} \\
gpt-3.5 & \cellcolor[HTML]{E1EEED}31.8 & \cellcolor[HTML]{DEECEC}32.8 & \upp{1.0} & \cellcolor[HTML]{D4E6E6}36.9 & \upp{5.1} & \cellcolor[HTML]{DFEDEC}32.4 & \upp{0.6} & \cellcolor[HTML]{E1EEED}31.8 & \equ{0.0} & \cellcolor[HTML]{E7F1F1}29.4 & \down{-2.4} \\
Copilot & \cellcolor[HTML]{ABCECD}53.4 & \cellcolor[HTML]{9FC7C7}58.0 & \upp{4.6} & \cellcolor[HTML]{8EBDBC}64.8 & \upp{11.4} & \cellcolor[HTML]{A2C9C8}57.1 & 3.7 & \cellcolor[HTML]{BDD9D8}46.2 & \down{-7.2} & \cellcolor[HTML]{BDD9D8}46.1 & \down{-7.3}

\\
\bottomrule
\end{tabular}
}
\end{table}

%% file: Tex/05-evaluation-rq4.tex
\subsection{RQ4. Could existing metrics distinguish contaminated/cleansed data?}

Fig.~\ref{fig:rq-score} shows six MIA-related scores on different code groups. The figure shows the results calculated on StarCoder; the results on other CLMs are similar, so we omit them due to space limits. 
Note that these scores indicate how ``natural'' the models found the inputs are. The lower the score,  the closer the input fits the models' training distribution.

\input{FigureTex/ppl_scores}

The contaminated/cleansed code groups generally share similar scores in all metrics. 
The mean perplexity of the code groups ranges from 3.37 (Code-2023) to 5.17 (Code-CodE). The largest score difference is observed in ppl\_lower, ranging from 4.06 (Code-2023) to 6.41 (Code-CodE). 
In addition, the Code-2023, Code-CodE, and the five refactored code groups are unseen for StarCoder whose cut-off date is March 2022, yet there is no apparent difference in these cleansed code groups compared with that on contaminated code groups. 

\begin{mdframed}[style=MyFrame]
\textit{\textbf{Finding 8:}} The existing six MIA-related metrics can \textbf{\textit{hardly distinguish the contaminated/cleansed}} data, so they may be hardly used for quantifying data contamination. 
\end{mdframed}

Surprisingly, though the recent code, \ie, Code-2023, is later than StarCoder's cut-off date (March 2022), all six metrics show absolute drops on it compared with other code groups.
Code-2023 fits CLMs' training distribution better than other code groups, so the model found it more natural (lower scores). 
This observation may partially explain the counterintuitiveness we observed in RQ1. CLMs find Code-2023 more natural than earlier code groups and closer to the training distribution; their performance will naturally be better.

One reason is the developers' tendency to duplicate codes through copy-paste.
However, this practice 
has been a long-established norm within the programming community, as discussed in prior literature~\cite{copypaste}. We would expect a similar pattern in previous data if it were the primary factor.
When we look at the earlier code groups (\eg, Code-2018 to Code-2021), the scores are stable without a significant drop, suggesting that other factors may influence this trend.

Another possible reason is the emergence of AI coding assistants such as Github Copilot.
As we mentioned in RQ1, nearly 40\% of the code has been written by GitHub Copilot in the last year. 
As this trend continues, more developers use AI as a programming aid, making the codes generated by AI more naturally fit the training distribution of the models.

In addition, looking at the refactored code groups in all six subplots, code refactoring appears not to have significantly impacted the model's familiarity with the code. The scores only exhibited minor disturbances. Among these refactoring operators, identifier renaming (Code-Renm) and performance measurement decoration (Code-Deco) have a larger impact, indicating that these two operators are more likely to result in a divergence from the original training data distribution. Such observation also echoes the findings in RQ3. 

\begin{mdframed}[style=MyFrame]
\textit{\textbf{Finding 9:}} Identifier renaming (\texttt{Renm}) and performance measurement decoration (\texttt{Deco}) have a greater impact on data, making CLMs unfamiliar with the data.
\\
\textbf{\textit{Implication:}} \texttt{Renm} and \texttt{Deco} may be more useful for alleviating data contamination.
\end{mdframed}

%% file: FigureTex/ppl_scores.tex
\begin{figure*}[th]
    \centering
    \includegraphics[width=1.0\textwidth]{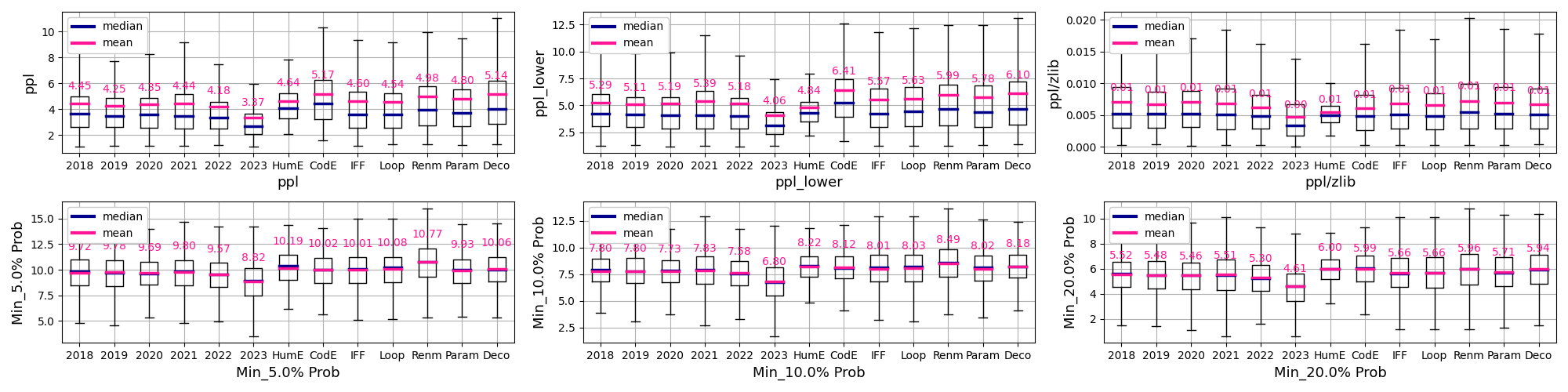}
\caption{Contamination-related Scores Over Various Code Groups}
\label{fig:rq-score}
\end{figure*} 

%% file: Tex/055-threat.tex
\section{Threats to validity}

We acknowledge several threats to the validity of our conclusions. First, \textbf{\textit{programming language selection bias}}. Our study confines itself to the analysis of Python, which, while being one of the most popular programming languages, does not represent other languages. We justify the selection of Python due to its widespread adoption and the belief that insights derived from popular languages may have broader relevance. 
Second, \textbf{\textit{coding task selection bias.}} The study focuses on code completion tasks with the masking of the last statement may not encapsulate the full range of capabilities required for other coding tasks. The performance of CLMs could vary on other tasks. We selected it because of its relevance to real-world coding practices and the advantage of minimizing human intervention, which can introduce additional variability. To enrich our understanding of CLMs' capabilities, we call for further research to explore a wider array of coding tasks. Third, \textit{\textbf{possible semantics overlapping between code groups.}} While efforts were made to prevent the overlap of code groups, the possibility of semantic overlap cannot be entirely eliminated. To mitigate this threat, three similarity metrics were employed to ascertain the distinctness of code groups, and the results suggest a relatively low level of similarity.

%% file: Tex/06-relatedwork.tex
\section{Related Work}\label{sec:related}

\subsection{Data Contamination in Large Language Models}
Several recent studies~\cite{sainz-etal-2023-nlp-contam,dataContamination2023,taskContamination2023,balloccu-etal-2024-leak,tirumala2022memorization,kandpal2022deduplicating,schick2020s,magar2022data} have explored data contamination in large language models (LLMs). 
Brown \etal~\cite{extracting21} analyzed the contamination in GPT-3, showing that the performance between contaminated and clean data is not necessarily different. 
Carlini \etal~\cite{tirumala2022memorization} found that the memorization of LLMs grows with model size, training data duplicates, and prompt length. 
Kandpal~\etal~\cite{kandpal2022deduplicating} identified that the success of privacy attacks on LLMs could be attributed to the sequence duplication in the training set. 
Razeghi~\etal~\cite{schick2020s} studied the correlations between LLMs' performance and the frequency of terms and observed that LLMs perform better on inputs with more frequent terms.  
Magar~\etal~\cite{magar2022data} studied the memorization and exploitation of masked language models in a controlled manner. 
They fine-tune two models, one with labeled test data and the other without. 
They identified the number of duplications of the contaminated data and the model size that affected model exploitation. 
Our work is based on different assumptions. 
Magar~\etal pre-trained and fine-tuned models (\ie, BERT) from scratch, with a controlled training set, aiming at identifying factors that affect model exploitation, 
while we study the efficacy of countermeasures for alleviating data contamination and identifying effective countermeasures.
\looseness=-1

\subsection{Membership Inference Attacks}
{{Membership inference attacks}} (MIAs) try to determine whether a particular data is contained in the model's training data. Various metrics are proposed to infer data membership, including LOSS~\cite{loss-mia}, reference models~\cite{extracting21}, perplexity~\cite{jelinek1977perplexity}, Zlib Entropy~\cite{gailly2004zlib},
Neighborhood attack~\cite{mattern2023membership}, Min-k\% Prob~\cite{topkmia2023},~\etc
Though MIAs are extensively studied in traditional deep learning models, 
the research on MIA in LLMs is limited. Recently, Duan~\etal~\cite{duan2024membership} conducted a large-scale MIA on LLMs and found that MIAs barely outperform random guessing across varying LLM sizes and domains. They also identified that different temporal ranges of data may affect model performance. These works aim at extracting sensitive data from LLMs, While our study considers how to evaluate CLMs fairly.

\subsection{Countermeasures Alleviating Data Contamination Threat}
Various benchmarks~\cite{jimenez2024swebench,mbpp2021,humaneval,evalplus,yu2023codereval,du2023classeval,VJBench23} for coding tasks are proposed to evaluate CLMs in diverse abilities and on a different scales. Most of these benchmarks were constructed or processed manually to ensure the quality of source code and other materials (\eg, docstrings, test cases). Yet, it is unclear how CLMs perform on these benchmarks compared with training data and how large the difference is. This study attempts to answer this question. 

Several works refactored code to alleviate data contamination. Wu~\etal~\cite{VJBench23} adopted identifier renaming and Code Structure Change to the code they collected. They conducted the refactoring manually, and they did not compare the performance change after code refactoring.
Besides, several studies~\cite{fan2023large,poldrack2023ai,noever2023chatbots} leverage code assistants such as GPT-4 to do the code refactoring. Yet, we avoid using AIs to code to get rid of reintroducing variants to the study.

%% file: Tex/07-conclusion.tex
\section{Conclusion}\label{sec:conclusion}
We conducted a systematic study on the data contamination threats of CLMs with contaminated and cleansed data. 
We investigated how CLMs perform on contaminated/recent code, code in curated benchmarks, and refactored code and found that CLMs in general achieve a stable performance between contaminated data and different types of cleansed data.
Our study indicates that existing countermeasures in alleviating data contamination threats may not be effective, calling for more discussion on the related topics of CLM-based software engineering research.

%% file: 01-main.bbl
\begin{thebibliography}{10}
\providecommand{\url}[1]{#1}
\csname url@samestyle\endcsname
\providecommand{\newblock}{\relax}
\providecommand{\bibinfo}[2]{#2}
\providecommand{\BIBentrySTDinterwordspacing}{\spaceskip=0pt\relax}
\providecommand{\BIBentryALTinterwordstretchfactor}{4}
\providecommand{\BIBentryALTinterwordspacing}{\spaceskip=\fontdimen2\font plus
\BIBentryALTinterwordstretchfactor\fontdimen3\font minus \fontdimen4\font\relax}
\providecommand{\BIBforeignlanguage}[2]{{%
\expandafter\ifx\csname l@#1\endcsname\relax
\typeout{** WARNING: IEEEtran.bst: No hyphenation pattern has been}%
\typeout{** loaded for the language `#1'. Using the pattern for}%
\typeout{** the default language instead.}%
\else
\language=\csname l@#1\endcsname
\fi
#2}}
\providecommand{\BIBdecl}{\relax}
\BIBdecl

\bibitem{deng2023large}
Y.~Deng, C.~S. Xia, H.~Peng, C.~Yang, and L.~Zhang, ``Large language models are zero-shot fuzzers: Fuzzing deep-learning libraries via large language models,'' in \emph{Proceedings of the 32nd ACM SIGSOFT international symposium on software testing and analysis}, 2023, pp. 423--435.

\bibitem{deng2024large}
Y.~Deng, C.~S. Xia, C.~Yang, S.~D. Zhang, S.~Yang, and L.~Zhang, ``Large language models are edge-case generators: Crafting unusual programs for fuzzing deep learning libraries,'' in \emph{Proceedings of the 46th IEEE/ACM International Conference on Software Engineering}, 2024, pp. 1--13.

\bibitem{xia2024fuzz4all}
C.~S. Xia, M.~Paltenghi, J.~Le~Tian, M.~Pradel, and L.~Zhang, ``Fuzz4all: Universal fuzzing with large language models,'' \emph{Proc. IEEE/ACM ICSE}, 2024.

\bibitem{xia2023automated}
C.~S. Xia, Y.~Wei, and L.~Zhang, ``Automated program repair in the era of large pre-trained language models,'' in \emph{2023 IEEE/ACM 45th International Conference on Software Engineering (ICSE)}.\hskip 1em plus 0.5em minus 0.4em\relax IEEE, 2023, pp. 1482--1494.

\bibitem{dataContamination2023}
\BIBentryALTinterwordspacing
S.~Golchin and M.~Surdeanu, ``Time travel in llms: Tracing data contamination in large language models,'' \emph{CoRR}, vol. abs/2308.08493, 2023. [Online]. Available: \url{https://doi.org/10.48550/arXiv.2308.08493}
\BIBentrySTDinterwordspacing

\bibitem{sainz-etal-2023-nlp-contam}
\BIBentryALTinterwordspacing
O.~Sainz, J.~Campos, I.~Garc{\'\i}a-Ferrero, J.~Etxaniz, O.~L. de~Lacalle, and E.~Agirre, ``{NLP} evaluation in trouble: On the need to measure {LLM} data contamination for each benchmark,'' in \emph{Findings of the Association for Computational Linguistics: EMNLP 2023}, H.~Bouamor, J.~Pino, and K.~Bali, Eds.\hskip 1em plus 0.5em minus 0.4em\relax Singapore: Association for Computational Linguistics, Dec. 2023, pp. 10\,776--10\,787. [Online]. Available: \url{https://aclanthology.org/2023.findings-emnlp.722}
\BIBentrySTDinterwordspacing

\bibitem{taskContamination2023}
\BIBentryALTinterwordspacing
C.~Li and J.~Flanigan, ``Task contamination: Language models may not be few-shot anymore,'' \emph{CoRR}, vol. abs/2312.16337, 2023. [Online]. Available: \url{https://doi.org/10.48550/arXiv.2312.16337}
\BIBentrySTDinterwordspacing

\bibitem{li2023open}
Y.~Li, ``An open source data contamination report for llama series models,'' \emph{arXiv preprint arXiv:2310.17589}, 2023.

\bibitem{VJBench23}
\BIBentryALTinterwordspacing
Y.~Wu, N.~Jiang, H.~V. Pham, T.~Lutellier, J.~Davis, L.~Tan, P.~Babkin, and S.~Shah, ``How effective are neural networks for fixing security vulnerabilities,'' in \emph{Proceedings of the 32nd ACM SIGSOFT International Symposium on Software Testing and Analysis}, ser. ISSTA 2023.\hskip 1em plus 0.5em minus 0.4em\relax New York, NY, USA: Association for Computing Machinery, 2023, p. 1282–1294. [Online]. Available: \url{https://doi.org/10.1145/3597926.3598135}
\BIBentrySTDinterwordspacing

\bibitem{li2023nuances}
T.-O. Li, W.~Zong, Y.~Wang, H.~Tian, Y.~Wang, S.-C. Cheung, and J.~Kramer, ``Nuances are the key: Unlocking chatgpt to find failure-inducing tests with differential prompting,'' in \emph{2023 38th IEEE/ACM International Conference on Automated Software Engineering (ASE)}.\hskip 1em plus 0.5em minus 0.4em\relax IEEE, 2023, pp. 14--26.

\bibitem{cao2023study}
J.~Cao, M.~Li, M.~Wen, and S.-c. Cheung, ``A study on prompt design, advantages and limitations of chatgpt for deep learning program repair,'' \emph{arXiv preprint arXiv:2304.08191}, 2023.

\bibitem{fan2023large}
A.~Fan, B.~Gokkaya, M.~Harman, M.~Lyubarskiy, S.~Sengupta, S.~Yoo, and J.~M. Zhang, ``Large language models for software engineering: Survey and open problems,'' \emph{arXiv preprint arXiv:2310.03533}, 2023.

\bibitem{tirumala2022memorization}
K.~Tirumala, A.~Markosyan, L.~Zettlemoyer, and A.~Aghajanyan, ``Memorization without overfitting: Analyzing the training dynamics of large language models,'' \emph{Advances in Neural Information Processing Systems}, vol.~35, pp. 38\,274--38\,290, 2022.

\bibitem{topkmia2023}
W.~Shi, A.~Ajith, M.~Xia, Y.~Huang, D.~Liu, T.~Blevins, D.~Chen, and L.~Zettlemoyer, ``Detecting pretraining data from large language models,'' \emph{arXiv preprint arXiv:2310.16789}, 2023.

\bibitem{extracting21}
\BIBentryALTinterwordspacing
N.~Carlini, F.~Tram{\`{e}}r, E.~Wallace, M.~Jagielski, A.~Herbert{-}Voss, K.~Lee, A.~Roberts, T.~B. Brown, D.~Song, {\'{U}}.~Erlingsson, A.~Oprea, and C.~Raffel, ``Extracting training data from large language models,'' in \emph{30th {USENIX} Security Symposium, {USENIX} Security 2021, August 11-13, 2021}, M.~D. Bailey and R.~Greenstadt, Eds.\hskip 1em plus 0.5em minus 0.4em\relax {USENIX} Association, 2021, pp. 2633--2650. [Online]. Available: \url{https://www.usenix.org/conference/usenixsecurity21/presentation/carlini-extracting}
\BIBentrySTDinterwordspacing

\bibitem{touvron2023llama}
H.~Touvron, L.~Martin, K.~Stone, P.~Albert, A.~Almahairi, Y.~Babaei, N.~Bashlykov, S.~Batra, P.~Bhargava, S.~Bhosale \emph{et~al.}, ``Llama 2: Open foundation and fine-tuned chat models,'' \emph{arXiv preprint arXiv:2307.09288}, 2023.

\bibitem{humaneval}
M.~Chen, J.~Tworek, H.~Jun, Q.~Yuan, H.~P. de~Oliveira~Pinto, J.~Kaplan, H.~Edwards, Y.~Burda, N.~Joseph, G.~Brockman, A.~Ray, R.~Puri, G.~Krueger, M.~Petrov, H.~Khlaaf, G.~Sastry, P.~Mishkin, B.~Chan, S.~Gray, N.~Ryder, M.~Pavlov, A.~Power, L.~Kaiser, M.~Bavarian, C.~Winter, P.~Tillet, F.~P. Such, D.~Cummings, M.~Plappert, F.~Chantzis, E.~Barnes, A.~Herbert-Voss, W.~H. Guss, A.~Nichol, A.~Paino, N.~Tezak, J.~Tang, I.~Babuschkin, S.~Balaji, S.~Jain, W.~Saunders, C.~Hesse, A.~N. Carr, J.~Leike, J.~Achiam, V.~Misra, E.~Morikawa, A.~Radford, M.~Knight, M.~Brundage, M.~Murati, K.~Mayer, P.~Welinder, B.~McGrew, D.~Amodei, S.~McCandlish, I.~Sutskever, and W.~Zaremba, ``Evaluating large language models trained on code,'' 2021.

\bibitem{mpbb2021}
J.~Austin, A.~Odena, M.~Nye, M.~Bosma, H.~Michalewski, D.~Dohan, E.~Jiang, C.~Cai, M.~Terry, Q.~Le \emph{et~al.}, ``Program synthesis with large language models,'' \emph{arXiv preprint arXiv:2108.07732}, 2021.

\bibitem{du2023classeval}
X.~Du, M.~Liu, K.~Wang, H.~Wang, J.~Liu, Y.~Chen, J.~Feng, C.~Sha, X.~Peng, and Y.~Lou, ``Classeval: A manually-crafted benchmark for evaluating llms on class-level code generation,'' 2023.

\bibitem{yu2023codereval}
H.~Yu, B.~Shen, D.~Ran, J.~Zhang, Q.~Zhang, Y.~Ma, G.~Liang, Y.~Li, T.~Xie, and Q.~Wang, ``Codereval: A benchmark of pragmatic code generation with generative pre-trained models,'' \emph{arXiv preprint arXiv:2302.00288}, 2023.

\bibitem{Lai2023DS1000}
Y.~Lai, C.~Li, Y.~Wang, T.~Zhang, R.~Zhong, L.~Zettlemoyer, W.-T. Yih, D.~Fried, S.~Wang, and T.~Yu, ``{DS}-1000: A natural and reliable benchmark for data science code generation,'' in \emph{Proceedings of the 40th International Conference on Machine Learning}, ser. Proceedings of Machine Learning Research, A.~Krause, E.~Brunskill, K.~Cho, B.~Engelhardt, S.~Sabato, and J.~Scarlett, Eds., vol. 202.\hskip 1em plus 0.5em minus 0.4em\relax PMLR, 23--29 Jul 2023, pp. 18\,319--18\,345.

\bibitem{jimenez2024swebench}
\BIBentryALTinterwordspacing
C.~E. Jimenez, J.~Yang, A.~Wettig, S.~Yao, K.~Pei, O.~Press, and K.~R. Narasimhan, ``{SWE}-bench: Can language models resolve real-world github issues?'' in \emph{The Twelfth International Conference on Learning Representations}, 2024. [Online]. Available: \url{https://openreview.net/forum?id=VTF8yNQM66}
\BIBentrySTDinterwordspacing

\bibitem{shirafuji2023refactoring}
A.~Shirafuji, Y.~Oda, J.~Suzuki, M.~Morishita, and Y.~Watanobe, ``Refactoring programs using large language models with few-shot examples,'' \emph{arXiv preprint arXiv:2311.11690}, 2023.

\bibitem{huang2023finbert}
A.~H. Huang, H.~Wang, and Y.~Yang, ``Finbert: A large language model for extracting information from financial text,'' \emph{Contemporary Accounting Research}, vol.~40, no.~2, pp. 806--841, 2023.

\bibitem{gruver2024large}
N.~Gruver, M.~Finzi, S.~Qiu, and A.~G. Wilson, ``Large language models are zero-shot time series forecasters,'' \emph{Advances in Neural Information Processing Systems}, vol.~36, 2024.

\bibitem{thestackv2}
``bigcode/the-stack-v2-dedup,'' \url{https://huggingface.co/datasets/bigcode/the-stack-v2-dedup}, 2023.

\bibitem{loss-mia}
S.~Yeom, I.~Giacomelli, M.~Fredrikson, and S.~Jha, ``Privacy risk in machine learning: Analyzing the connection to overfitting,'' in \emph{2018 IEEE 31st Computer Security Foundations Symposium (CSF)}, 2018, pp. 268--282.

\bibitem{kandpal2022deduplicating}
N.~Kandpal, E.~Wallace, and C.~Raffel, ``Deduplicating training data mitigates privacy risks in language models,'' in \emph{International Conference on Machine Learning}.\hskip 1em plus 0.5em minus 0.4em\relax PMLR, 2022, pp. 10\,697--10\,707.

\bibitem{jelinek1977perplexity}
F.~Jelinek, R.~L. Mercer, L.~R. Bahl, and J.~K. Baker, ``Perplexity—a measure of the difficulty of speech recognition tasks,'' \emph{The Journal of the Acoustical Society of America}, vol.~62, no.~S1, pp. S63--S63, 1977.

\bibitem{duan2024membership}
M.~Duan, A.~Suri, N.~Mireshghallah, S.~Min, W.~Shi, L.~Zettlemoyer, Y.~Tsvetkov, Y.~Choi, D.~Evans, and H.~Hajishirzi, ``Do membership inference attacks work on large language models?'' \emph{arXiv preprint arXiv:2402.07841}, 2024.

\bibitem{mbpp2021}
J.~Austin, A.~Odena, M.~Nye, M.~Bosma, H.~Michalewski, D.~Dohan, E.~Jiang, C.~Cai, M.~Terry, Q.~Le \emph{et~al.}, ``Program synthesis with large language models,'' \emph{arXiv preprint arXiv:2108.07732}, 2021.

\bibitem{humanevaldata}
``openai/human-eval,'' \url{https://github.com/openai/human-eval/commits/master/data}, 2021.

\bibitem{codereval4python}
``Codereval/codereval,'' \url{https://github.com/CoderEval/CoderEval/commits/main/CoderEval4Python.json}, 2023.

\bibitem{baqais2020automatic}
A.~A.~B. Baqais and M.~Alshayeb, ``Automatic software refactoring: a systematic literature review,'' \emph{Software Quality Journal}, vol.~28, no.~2, pp. 459--502, 2020.

\bibitem{al2017empirical}
J.~Al~Dallal and A.~Abdin, ``Empirical evaluation of the impact of object-oriented code refactoring on quality attributes: A systematic literature review,'' \emph{IEEE Transactions on Software Engineering}, vol.~44, no.~1, pp. 44--69, 2017.

\bibitem{carlini2019secret}
N.~Carlini, C.~Liu, {\'U}.~Erlingsson, J.~Kos, and D.~Song, ``The secret sharer: Evaluating and testing unintended memorization in neural networks,'' in \emph{28th USENIX Security Symposium (USENIX Security 19)}, 2019, pp. 267--284.

\bibitem{li2023estimating}
Y.~Li, ``Estimating contamination via perplexity: Quantifying memorisation in language model evaluation,'' \emph{arXiv preprint arXiv:2309.10677}, 2023.

\bibitem{raychev2014code}
V.~Raychev, M.~Vechev, and E.~Yahav, ``Code completion with statistical language models,'' in \emph{Proceedings of the 35th ACM SIGPLAN conference on programming language design and implementation}, 2014, pp. 419--428.

\bibitem{svyatkovskiy2019pythia}
A.~Svyatkovskiy, Y.~Zhao, S.~Fu, and N.~Sundaresan, ``Pythia: Ai-assisted code completion system,'' in \emph{Proceedings of the 25th ACM SIGKDD international conference on knowledge discovery \& data mining}, 2019, pp. 2727--2735.

\bibitem{codeComplexity}
P.~Oman and J.~Hagemeister, ``Metrics for assessing a software system's maintainability,'' in \emph{Proceedings Conference on Software Maintenance 1992}, 1992, pp. 337--344.

\bibitem{codeMetrics}
D.~Coleman, D.~Ash, B.~Lowther, and P.~Oman, ``Using metrics to evaluate software system maintainability,'' \emph{Computer}, vol.~27, no.~8, pp. 44--49, 1994.

\bibitem{radon}
\url{https://radon.readthedocs.io/en/latest/intro.html#cyclomatic-complexity}, 2012.

\bibitem{bieri1955cognitive}
J.~Bieri, ``Cognitive complexity-simplicity and predictive behavior.'' \emph{The Journal of Abnormal and Social Psychology}, vol.~51, no.~2, p. 263, 1955.

\bibitem{cog}
\url{https://pypi.org/project/cognitive-complexity/}, 2022.

\bibitem{codebleu}
\BIBentryALTinterwordspacing
S.~Ren, D.~Guo, S.~Lu, L.~Zhou, S.~Liu, D.~Tang, N.~Sundaresan, M.~Zhou, A.~Blanco, and S.~Ma, ``Codebleu: a method for automatic evaluation of code synthesis,'' \emph{CoRR}, vol. abs/2009.10297, 2020. [Online]. Available: \url{https://arxiv.org/abs/2009.10297}
\BIBentrySTDinterwordspacing

\bibitem{neardup}
\BIBentryALTinterwordspacing
K.~Lee, D.~Ippolito, A.~Nystrom, C.~Zhang, D.~Eck, C.~Callison{-}Burch, and N.~Carlini, ``Deduplicating training data makes language models better,'' \emph{CoRR}, vol. abs/2107.06499, 2021. [Online]. Available: \url{https://arxiv.org/abs/2107.06499}
\BIBentrySTDinterwordspacing

\bibitem{minhash}
A.~Z. Broder, ``Identifying and filtering near-duplicate documents,'' in \emph{Combinatorial Pattern Matching}, R.~Giancarlo and D.~Sankoff, Eds.\hskip 1em plus 0.5em minus 0.4em\relax Berlin, Heidelberg: Springer Berlin Heidelberg, 2000, pp. 1--10.

\bibitem{wordhoard}
\url{https://github.com/johnbumgarner/wordhoard}, 2024.

\bibitem{gailly2004zlib}
J.-l. Gailly and M.~Adler, ``Zlib compression library,'' 2004.

\bibitem{zhang2023ethicist}
Z.~Zhang, J.~Wen, and M.~Huang, ``Ethicist: Targeted training data extraction through loss smoothed soft prompting and calibrated confidence estimation,'' \emph{arXiv preprint arXiv:2307.04401}, 2023.

\bibitem{Kocetkov2022TheStack}
D.~Kocetkov, R.~Li, L.~Ben~Allal, J.~Li, C.~Mou, C.~Muñoz~Ferrandis, Y.~Jernite, M.~Mitchell, S.~Hughes, T.~Wolf, D.~Bahdanau, L.~von Werra, and H.~de~Vries, ``The stack: 3 tb of permissively licensed source code,'' \emph{Preprint}, 2022.

\bibitem{starcoder}
``bigcode/starcoder,'' \url{https://huggingface.co/bigcode/starcoder}, 2023.

\bibitem{starcoderbase}
``bigcode/starcoderbase,'' \url{https://huggingface.co/bigcode/starcoderbase}, 2023.

\bibitem{Tunstall2023starchat-alpha}
L.~Tunstall, N.~Lambert, N.~Rajani, E.~Beeching, T.~Le~Scao, L.~von Werra, S.~Han, P.~Schmid, and A.~Rush, ``Creating a coding assistant with starcoder,'' \emph{Hugging Face Blog}, 2023, https://huggingface.co/blog/starchat.

\bibitem{starchatbeta}
``Huggingfaceh4/starchat-beta,'' \url{https://huggingface.co/HuggingFaceH4/starchat-beta/commit/4d8424c69643590f193cc97dc7eebff66500ebc6}, 2023.

\bibitem{openassistant}
``timdettmers/openassistant-guanaco,'' \url{https://huggingface.co/datasets/timdettmers/openassistant-guanaco}, 2023.

\bibitem{oasst1}
``Openassistant/oasst1,'' \url{https://huggingface.co/datasets/OpenAssistant/oasst1/tree/main}, 2023.

\bibitem{wizardCoder23}
\BIBentryALTinterwordspacing
Z.~Luo, C.~Xu, P.~Zhao, Q.~Sun, X.~Geng, W.~Hu, C.~Tao, J.~Ma, Q.~Lin, and D.~Jiang, ``Wizardcoder: Empowering code large language models with evol-instruct,'' \emph{CoRR}, vol. abs/2306.08568, 2023. [Online]. Available: \url{https://doi.org/10.48550/arXiv.2306.08568}
\BIBentrySTDinterwordspacing

\bibitem{wizardcoder}
``Wizardlm/wizardcoder,'' \url{https://huggingface.co/WizardLM/WizardCoder-15B-V1.0}, 2023.

\bibitem{codealpaca}
S.~Chaudhary, ``Code alpaca: An instruction-following llama model for code generation,'' \url{https://github.com/sahil280114/codealpaca}, 2023.

\bibitem{codealpacacommit}
``sahil280114/codealpaca,'' \url{https://github.com/sahil280114/codealpaca/commit/d269da106a579a623a654529b3cb91b5dfa9c72f}, 2023.

\bibitem{codellama7b}
``codellama/codellama-7b-instruct-hf,'' \url{https://huggingface.co/codellama/CodeLlama-7b-Instruct-hf}, 2023.

\bibitem{codellama7binstructcommit}
``codellama/codellama-7b-instruct-hf commit,'' \url{https://huggingface.co/codellama/CodeLlama-7b-Instruct-hf/commit/65db8fcae13921e49c3dd0a2be4757102d0e723f}, 2023.

\bibitem{llama27b}
``meta-llama/llama-2-7b,'' \url{https://huggingface.co/meta-llama/Llama-2-7b}, 2023.

\bibitem{phindcodellama34bv2}
``Phind/phind-codellama-34b-v2,'' \url{https://huggingface.co/Phind/Phind-CodeLlama-34B-v2}, 2023.

\bibitem{phind}
``Phind/phind-codellama-34b-v2,'' \url{https://huggingface.co/Phind/Phind-CodeLlama-34B-v2/commit/29c3be6006297754f344ba05678c038b0b77f6c0}, 2023.

\bibitem{du2022glm}
Z.~Du, Y.~Qian, X.~Liu, M.~Ding, J.~Qiu, Z.~Yang, and J.~Tang, ``Glm: General language model pretraining with autoregressive blank infilling,'' in \emph{Proceedings of the 60th Annual Meeting of the Association for Computational Linguistics (Volume 1: Long Papers)}, 2022, pp. 320--335.

\bibitem{zeng2022glm}
A.~Zeng, X.~Liu, Z.~Du, Z.~Wang, H.~Lai, M.~Ding, Z.~Yang, Y.~Xu, W.~Zheng, X.~Xia \emph{et~al.}, ``Glm-130b: An open bilingual pre-trained model,'' \emph{arXiv preprint arXiv:2210.02414}, 2022.

\bibitem{chatglm}
``Thudm/chatglm3-6b,'' \url{https://huggingface.co/THUDM/chatglm3-6b}, 2023.

\bibitem{chatglm36b}
``Thudm/chatglm3-6b,'' \url{https://huggingface.co/THUDM/chatglm3-6b/commit/62acdaa77a5742d120acf9b6419656b403218c3d}, 2023.

\bibitem{gpt35turbo}
``Gpt-3.5-turbo model availability,'' \url{https://learn.microsoft.com/en-us/azure/ai-services/openai/concepts/models\#gpt-35-turbo-model-availability}, 2023.

\bibitem{githubcopilot}
``Github copilot,'' \url{https://copilot.microsoft.com/}, 2023.

\bibitem{githubcopilotintroduce}
``Introducing github copilot: Ai pair programmer,'' \url{https://github.blog/2021-06-29-introducing-github-copilot-ai-pair-programmer/}, 2023.

\bibitem{githubcopilotupdate}
``Github copilot november 30th update,'' \url{https://github.blog/changelog/2023-11-30-github-copilot-november-30th-update/}, 2023.

\bibitem{gpt4turbo}
``Gpt-4 and gpt-4-turbo preview,'' \url{https://learn.microsoft.com/en-us/azure/ai-services/openai/concepts/models\#gpt-4-and-gpt-4-turbo-preview}, 2023.

\bibitem{nuclear}
\BIBentryALTinterwordspacing
A.~Holtzman, J.~Buys, L.~Du, M.~Forbes, and Y.~Choi, ``The curious case of neural text degeneration,'' in \emph{8th International Conference on Learning Representations, {ICLR} 2020, Addis Ababa, Ethiopia, April 26-30, 2020}.\hskip 1em plus 0.5em minus 0.4em\relax OpenReview.net, 2020. [Online]. Available: \url{https://openreview.net/forum?id=rygGQyrFvH}
\BIBentrySTDinterwordspacing

\bibitem{ouyang2023llm}
S.~Ouyang, J.~M. Zhang, M.~Harman, and M.~Wang, ``Llm is like a box of chocolates: the non-determinism of chatgpt in code generation,'' \emph{arXiv preprint arXiv:2308.02828}, 2023.

\bibitem{githubcopilotavailable}
``Github copilot is generally available to all developers,'' \url{https://github.blog/2022-06-21-github-copilot-is-generally-available-to-\\all-developers/}, 2022.

\bibitem{copypaste}
M.~Kim, L.~Bergman, T.~Lau, and D.~Notkin, ``An ethnographic study of copy and paste programming practices in oopl,'' in \emph{Proceedings. 2004 International Symposium on Empirical Software Engineering, 2004. ISESE '04.}, 2004, pp. 83--92.

\bibitem{balloccu-etal-2024-leak}
\BIBentryALTinterwordspacing
S.~Balloccu, P.~Schmidtov{\'a}, M.~Lango, and O.~Dusek, ``Leak, cheat, repeat: Data contamination and evaluation malpractices in closed-source {LLM}s,'' in \emph{Proceedings of the 18th Conference of the European Chapter of the Association for Computational Linguistics (Volume 1: Long Papers)}, Y.~Graham and M.~Purver, Eds.\hskip 1em plus 0.5em minus 0.4em\relax St. Julian{'}s, Malta: Association for Computational Linguistics, Mar. 2024, pp. 67--93. [Online]. Available: \url{https://aclanthology.org/2024.eacl-long.5}
\BIBentrySTDinterwordspacing

\bibitem{schick2020s}
T.~Schick and H.~Sch{\"u}tze, ``It's not just size that matters: Small language models are also few-shot learners,'' \emph{arXiv preprint arXiv:2009.07118}, 2020.

\bibitem{magar2022data}
I.~Magar and R.~Schwartz, ``Data contamination: From memorization to exploitation,'' \emph{arXiv preprint arXiv:2203.08242}, 2022.

\bibitem{mattern2023membership}
J.~Mattern, F.~Mireshghallah, Z.~Jin, B.~Sch{\"o}lkopf, M.~Sachan, and T.~Berg-Kirkpatrick, ``Membership inference attacks against language models via neighbourhood comparison,'' \emph{arXiv preprint arXiv:2305.18462}, 2023.

\bibitem{evalplus}
\BIBentryALTinterwordspacing
J.~Liu, C.~S. Xia, Y.~Wang, and L.~Zhang, ``Is your code generated by chat{GPT} really correct? rigorous evaluation of large language models for code generation,'' in \emph{Thirty-seventh Conference on Neural Information Processing Systems}, 2023. [Online]. Available: \url{https://openreview.net/forum?id=1qvx610Cu7}
\BIBentrySTDinterwordspacing

\bibitem{poldrack2023ai}
R.~A. Poldrack, T.~Lu, and G.~Begu{\v{s}}, ``Ai-assisted coding: Experiments with gpt-4,'' \emph{arXiv preprint arXiv:2304.13187}, 2023.

\bibitem{noever2023chatbots}
D.~Noever and K.~Williams, ``Chatbots as fluent polyglots: Revisiting breakthrough code snippets,'' \emph{arXiv preprint arXiv:2301.03373}, 2023.

\end{thebibliography}
